\documentclass[letterpaper]{article}
\usepackage{jheppub}
\pdfoutput=1
\usepackage{color,latexsym, array,multirow,  verbatim}
\usepackage[config, singlelinecheck=true]{caption,subfig}
\usepackage{url}
\usepackage{multirow}

\newcommand{\gev}{~\text{GeV}}

\newcommand{\z}{z}
\newcommand{\f}{\rho}

\newcommand{\dB}{\delta B}

\def\simgt{\mathrel{\lower2.5pt\vbox{\lineskip=0pt\baselineskip=0pt
           \hbox{$>$}\hbox{$\sim$}}}}
\def\simlt{\mathrel{\lower2.5pt\vbox{\lineskip=0pt\baselineskip=0pt
           \hbox{$<$}\hbox{$\sim$}}}}
  \title{Jet Sampling: Improving Event Reconstruction through Multiple Interpretations}

\author{Dilani Kahawala,}
\author{David Krohn,}
\author{and Matthew D.~Schwartz}

\affiliation{Department of Physics, Harvard University, Cambridge MA, 02138}

\emailAdd{kahawala@physics.harvard.edu}
\emailAdd{dkrohn@physics.harvard.edu}
\emailAdd{schwartz@physics.harvard.edu}

\abstract{The classification of events involving jets as signal-like or background-like can depend strongly
on the jet algorithm used and its parameters. 
This is partly due to the fact that standard jet algorithms yield a single partition of the particles
in an event into jets, even if no particular choice stands out from the others.
As an alternative, we propose that one should consider multiple interpretations of each event,
generalizing the Qjets procedure to event-level analysis. With multiple interpretations,
an event is no longer restricted to either satisfy cuts or not satisfy them -- it can be assigned
a weight between 0 and 1 based on how well it satisfies the cuts. These cut-weights can then be used
to improve the discrimination power of an analysis or reduce the uncertainty on mass or cross-section
measurements. For example, using this approach on a Higgs plus $Z$ boson sample, with $H\to b\bar{b}$
we find an 28\% improvement in significance can be realized at the 8 TeV LHC.
Through a number of other examples, we show various ways in which having multiple interpretations can be useful on the event level.
}

\newcommand{\beq}{\begin{equation}}% can be used as {equation} or {eqnarray}
\newcommand{\eeq}{\end{equation}}
\newcommand{\beqs}{\begin{eqnarray}}% can be used as {equation} or {eqnarray}
\newcommand{\eeqs}{\end{eqnarray}}

\newcommand{\met}{\ensuremath{{\not\mathrel{E}}_T}}

\preprint{}

\begin{document}

\maketitle

\section{Introduction}

Almost every event recorded at the Large Hadron Collider contains some number of jets. 
Sometimes the jets are the objects of interest, as in a search for dijet resonances. Sometimes they
are indications of contamination and a jet veto can be used to increase
signal purity. Even in events that are predominantly electroweak some amount of jet
activity is usually present. Techniques for analyzing jets, in particular, the substructure
of jets, have been increasing in sophistication in recent years. Some recent reviews are~\cite{Salam:2009jx,Altheimer:2012mn, Shelton:2013an,Plehn:2011tg}.

To use jets for any sort of analysis, one first needs a way to  translate the hadronic activity in the event into a set of jets. 
 At the LHC, this is done almost universally with sequential recombination algorithms.
 These algorithms, such as the  anti-$k_T$~\cite{Cacciari:2008gp},
 Cambridge/Aachen~\cite{Dokshitzer:1997in,Wobisch:1998wt}, 
and $k_T$~\cite{Catani:1993hr,Ellis:1993tq} algorithms,
 assemble jets by merging particles in a sequence determined by some fixed distance measure.
 The result of applying such an  algorithm to an event is a tree containing a sequence of branchings.  The jets resulting from running a jet algorithm represent the algorithm's best guess as to which particles 
should be associated with the fragmentation of the same hard parton. 
In this paper (unlike in~\cite{Ellis:2012sn}) we will only be interested in which particles end up in which jet, not the structure of the
clustering tree.

In the majority of cases, such as when there are a few, well-separated jets,
 the best guess interpretation from any algorithm provides an excellent representation of the event.
However, for events with multiple and overlapping jets, the interpretations can differ greatly among
algorithms, or even when the parameters (such as the jet size $R$) of a single algorithm are varied.
  Ideally, one would like to treat events which are sensitive to the jet algorithm or jet parameters differently from ones which are more robust to algorithmic variations.
 In this paper, we propose a way to consider multiple interpretations of an event at once.

Intuitively it makes sense that considering multiple interpretations of an event should yield useful information. 
Indeed, probabilistic jet algorithms were first discussed long ago in relation to improving the behavior of seeded jet algorithms~\cite{Giele:1997ac}.
 Other related  approaches to jets include combining observables to improve discovery significance~\cite{Soper:2010xk,Cui:2010km,Gallicchio:2010dq}, 
comparing multiple interpretations of jet reconstruction with models of showering in signal and background processes~\cite{Soper:2011cr,Soper:2012pb}, 
and measuring the ``fuzziness'' of jet reconstruction~\cite{Volobouev:2011zz}. 
 Here we consider how multiple interpretations of each event can be used to turn single observables (e.g. dijet invariant mass) into distributions.
 This idea was proposed in~\cite{Ellis:2012sn} and called Qjets. In~\cite{Ellis:2012sn}
a proof-of-concept application of Qjets was given which focused on tree-based jet substructure. 
It was shown that Qjets can improve the statistical discriminating power in the search for boosted hadronically decaying objects.
 In this paper, we apply the multiple-interpretations aspect of the Qjets approach to jet reconstruction over a full event.
  
The basic idea behind Qjets is to sample interpretations near what a traditional jet algorithm would give. 
During a clustering step, a traditional jet algorithm merges the two closest particles based on some distance measure.
One possible way to sample interpretations around this standard interpretation is, rather than always merging the two closest particles, 
to merge two particles with some probability depending on how close they are.
The result is a set of $N$ interpretations of each event.

There are a number of ways one can process these $N$ interpretations. 
In~\cite{Ellis:2012sn}, the $N$ trees constructed from the particles in a single jet were pruned~\cite{Ellis:2009me}. Pruning throws out some particles based on the branching sequence in the tree. Since the pruned trees have different particles for each tree, the jet properties are different. For example,
since $N$ different jet masses result one can look at the width of the mass distribution for a single jet. This width, called volatility in~\cite{Ellis:2012sn}, was shown to be a useful discriminant between signal and background jets in certain cases. 

In this paper, we apply the multiple interpretations idea of Qjets to an entire event, and we do not apply pruning (or any other grooming procedure). Instead, we exploit the fact that different clusterings
will give jets with different 4-vectors. For example, if a particle is halfway between two jets, it might get clustered into each jet half the time. 
Or a particle which classical anti-$k_T$ clusters with the beam now has some probability to be clustered into a jet.
With multiple interpretations, particles can be associated with many different jets, in contrast to classical algorithms where 
each particle is always associated with exactly one jet.
%Also, using Qjets particles which would get clustered into a beam classical can sometimes get clustered into jets using Qjets. 
%In this way, we can  a more complete picture of an event. 

The result of applying Qjets to an event is a set of $N$ interpretations of that event. One way to process these interpretations is to apply some cuts to them, as one would in a classical analysis. For example,
one can impose a dijet mass window cut or a $p_T$ cut. While with a classical algorithm, an event would either pass or not pass the cuts, with Qjets, a fraction $\z$ of the interpretations pass. We call $\z$ the {\bf{cut-weight}}. Events with $\z$ close to 1 are then very likely to be signal, while events with $\z$ close to zero are unlikely to be signal. 
Although one can try cutting directly on $\z$ (similar to cuts on volatility in~\cite{Ellis:2012sn}),
it is better to use $\z$ to compute a statistical weight for a given event. 
That is, instead of throwing events out, each event is weighted by how signal-like it appears.
Then one simply constructs the distribution of, say, the dijet invariant mass, with each event
weighted by its $\z$-value. The statistical fluctuations on this weighted invariant mass will be smaller (often much smaller) than if $\z=0$ or $\z=1$
are the only possibilities considered (as in a classical analysis).
We will show that using weighted events in this way can provide significant improvements in 
the size of a signal divided by the characteristic background uncertainty, $S/\dB$, for many event classes. 

In this paper we consider 4 processes: 1)  $Z+H$ with $H\to b\bar{b}$, 2) a heavy scalar $\phi$ produced in association with a $Z$ boson with $\phi \to$~dijets,  3) 1 TeV dijet resonance, and 4) a heavy scalar decaying to 2 other scalars which each decay to dijets.
In cases where the event topology is simple and unambiguous, for example when there are two well separated jets, we find that standard algorithms perform quite well and the use of multiple interpretations only provides a marginal improvement. However, in more complex cases where events have jets with potentially overlapping boundaries, 
using the multiple interpretations can substantially improve significance over standard cut-based analyses. As an important example, we find a 28\% improvement in $S/\dB$ for $Z+H$ over its $Z+b\bar{b}$ irreducible background.

This article is structured as follows.  In Sec.~\ref{sec:alg} we will present a modification of the anti-$k_T$ algorithm to make it non-deterministic. Modifying the jet algorithm in this way generates a Monte-Carlo sampling of the distribution of interpretations around the best-guess interpretation.
Some ways to visualize the effect of multiple interpretations are presented in Sec.~\ref{sec:jetarea} and~\ref{sec:frac}.
In Sec.~\ref{sec:stats} we derive a formula for the statistical significance using our method.
Sec.~\ref{sec:demo} applies the algorithm to several samples of phenomenological interest.
Some comments on the speed of the algorithm are given in Sec.~\ref{sec:speed}.
Conclusions are  in Sec.~\ref{sec:conc}. 

\section{Qanti-$k_T$: a non-deterministic anti-$k_T$ algorithm}\label{sec:alg}
We begin by describing how the anti-$k_T$  algorithm can be modified to provide multiple interpretations of an event. 
While one would ideally  sample every possible reconstruction of an event, 
collider events typically contain a large number of final state particles so this is impractical. Instead we generate a representative sample of interpretations by using a Monte-Carlo integration type approach.
 A fastjet plugin with and implementation of this Qanti-$k_T$ algorithm is available at \url{http://jets.physics.harvard.edu/Qantikt}.
% 
% 
% \section {Algorithm}
% Below we will present a detained prescription to modify a sequential recombination algorithm, focusing on anti-$k_T$, so that it samples other similar clustering histories.  We will also discuss the algorithm's parameters and some tricks to improve its speed.  
% 
% \subsection{Procedure}

%assigning each of these a weight proportional to the likelihood the interpretation came from a signal process,  

The Qanti-$k_T$ algorithm works as follows. The input is a set of 4-vectors representing each particle's 4-momenta. These can be the stable hadrons in an event, charged tracks coming from a primary interaction, calorimeter cells, topoclusters, or the output of a Monte Carlo.

\begin{enumerate}
\item First calculate the distances $d_{ij}$ between each pair of 4-vectors and also the distances $d_{iB}$ between each 4-vector and
the beam. 
% 
% of the vectors (jet-jet distances $d_{ij}$)
% We begin by calculating all the distances between each of the vectors (jet-jet distances $d_{ij}$) and also between each vector and the beam (jet-beam distances $d_{iB}$). 
The metric used for the distance calculation is that of anti-$k_T$, although the procedure can easily be modified to work with C/A or $k_T$. The anti-$k_T$ distance measure is
  \begin{equation}
  \label{eq:dij}
   d_{ij} = {\rm min}\left( p_{Ti}^{-2}, p_{Tj}^{-2} \right) \frac{\Delta R_{ij}^2}{R^2},
  \end{equation}
and  
  \begin{equation}
   d_{iB} = p_{Ti}^{-2},
  \end{equation}
where $\Delta R_{ij} = \sqrt{(y_i - y_j)^2 + (\phi_i - \phi_j)^2}$ is the angular distance between a pair of 4-momenta $i$ and $j$ with
$y$ the rapidity and $\phi$ the azimuthal angle. $R$ is a free parameter in the anti-$k_T$ algorithm, representing the size of the final jets
one is interested in.

\item A weight is then assigned to each pair:
  \begin{equation}
  \label{eq:omega}
   \omega_{ij}^{(\alpha)} = {\rm exp}\left\{  -\alpha \frac{(d_{ij}- d^{\rm min})}{d^{\rm min}} \right\},
  \end{equation}
where $d^{\rm min}$ is the minimum distance over all pairs at this stage of the clustering and $\alpha$ is a real number called 
{\bf rigidity} in~\cite{Ellis:2012sn}.

\item A random number is used to choose a pair to merge. The probability of merging a given pair is
  \begin{equation}
  P_{ij}^{(\alpha)} = \frac{\omega_{ij}}{\sum_{i,j} \omega_{ij} }
  \end{equation}

\item Repeat until all particles have been merged into jets or a beam.
\end{enumerate}

 At its heart, the Qanti-$k_T$ algorithm is still a sequential recombination algorithm. However,
the weights and their Monte-Carlo sampling modify the order of the merging and change which particles get clustered into each jet or the beam
on each iteration.
%the essential difference is in the way we modify the order of merging by introducing a set of weights and an element of randomness.
 In a traditional sequential recombination algorithm the jets closest in distance  are merged first. In Qanti-$k_T$, jet-jet  or jet-beam distance is assigned a weight, controlled by a parameter $\alpha$, which allows the recombination order to vary. 
 For a given event we  find it is typically sufficient to repeat the Qanti-$k_T$ procedure a few tens of times at the same value of $\alpha$ for our results to stabilize\footnote{In this paper we will always run Qanti-$k_T$ $N=100$ times per event.}. 
The result can be thought of as a Monte-Carlo calculation of the distribution of interpretations around a best guess.

When $\alpha = 0$, all distances are given equal weighting, which means that particles far apart could be merged into the same jet early in the clustering. Somewhat surprisingly, despite the random clustering, in~\cite{Ellis:2012sn} it was found that Qanti-$k_T$ can still distinguish 
signal from background even when $\alpha=0$, although that will not be the case here. 
As $\alpha$ increases in value clusterings closer to those of anti-$k_T$ have higher weights and are consequently more likely to be realized.
One can think of $\alpha$ somewhat like $\frac{1}{\hbar}$. In analogy with with the $\hbar\rightarrow 0$ limit of quantum mechanics we term the  $\alpha \rightarrow \infty$ limit the {\it classical limit}.  In the classical limit
%the behavior of the classical anti-$k_T$ algorithm, 
the pair of particles closest in distance is always merged and the diversity of interpretations is lost.
% as the large weighting for short distances mean that the same particles get merged in every interpretation. 

In addition to $\alpha$, the jet radius parameter $R$ can also be varied. For finite $\alpha$ the final jets are no longer circles of radius $R$ as they are in classical anti-$k_T$. Indeed, with Qanti-$k_T$, there is no longer even a precise notion of where the jets are. This can be seen in Fig.~\ref{fig:coloredplots} below. As a result, there is less sensitivity to the precise choice of $R$ when using Qanti-$k_T$ than when using classical algorithms. This speaks to the general trends observed in~\cite{Ellis:2012sn}: with Qjets results depend much more weakly on the jet algorithm and algorithm parameters than with classical jets.

It is worth pointing out that the Qanti-$k_T$ algorithm is  infrared and collinear (IRC) safe. IRC safety means that when an arbitrarily soft particle is added or a particle is split
into two particles in the same direction, the results are unchanged.
% The basic idea is that infinitely soft and/or collinear emissions in QCD are incalculable, and so their presence or absence should not affect the results of a jet algorithm.  More formally, for an algorithm to be IRC safe its output must remain unchanged when an emission is split into two collinear pieces or whenever an infinitesimally soft emission is added or subtracted anywhere in the detector.
Qanti-$k_T$ is IRC safe as long as $\alpha>0$.  To see this, first note that all sequential recombination algorithms are by their nature infrared safe -- any infinitesimally soft emission will simply be clustered with harder radiation during the recombination process and will thus have no effect on the final
 outcome. For collinear splittings, one might worry that when non-determinism is added to the clustering, if a particle is split in two, the two halves might be clustered differently.   However,
 note that $d_{ij}\propto (\Delta R)^2$ (see Eq.~\ref{eq:dij}), and therefore  $d_{ij}$ between 
two particles which are exactly collinear will be exactly  zero.
When this happens, $d^\text{min}=0$ as well and so, for $\alpha>0$, $P_{ij}^\alpha=1$ for collinear particles and $P_{ij}^\alpha=0$ otherwise. Thus, collinear particles will always be clustered before non-collinear ones, and collinear particles will always end up in the same jet. 

% 
% In the limit of $\alpha\rightarrow\infty$ this controls the size of the jets, and so for smaller $\alpha$ we can roughly think of $R$ as governing the size of the jets.
% 
%  As a result, the shape of the jet may deviate significantly from the circular jets we usually observe using
% anti-$k_T$. Due to the use of random numbers in deciding which four-momenta get merged, running the same algorithm multiple times on the same event gives us multiple different interpretations of the final set of jets.
%  The particles that get clustered into a particular jet will vary from interpretation to interpretation. As $\alpha$ increases in value clusterings closer to those of anti-$k_T$ are given a higher weighting and are consequently more likely to be realized. In analogy with with the $\hbar\rightarrow 0$ limit of quantum mechanics we term the  $\alpha \rightarrow \infty$ limit the {\it classical limit}.  Here we regain the behavior of the classical anti-$k_T$ algorithm, where the pair of particles closest in distance is merged 
% first. At the same time, the diversity of interpretations is lost as the large weighting for short distances mean that the same particles get merged in every interpretation. 

\section{Overlapping jets and jet area}\label{sec:jetarea}
Before applying Qanti-$k_T$ to a signal/background discrimination task, we can explore how it differs from classical algorithms.
An advantage of Qanti-$k_T$ is that particles are not always clustered into the same jets. This is particularly useful in contexts where jets overlap. With overlapping jets, classical algorithms must assign each particle to exactly one jet. But Qanti-$k_T$ can split the particles into each jet some fraction of the time.\footnote{A note on our sample composition: we generate our signal and background events using a combination of Madgraph v5.7~\cite{Alwall:2011uj} and Pythia v6.4~\cite{Sjostrand:2006za}.  All events were generated assuming a 8 TeV LHC. We group the visible output of Pythia into massless $\delta \eta \times \delta \phi = 0.1 \times 0.1$ massless  cells with $|\eta| < 5$. Each type of event is analyzed with both Qanti-$k_T$ and also standard anti-$k_T$  for comparison. We use Fastjet v2.4.2~\cite{Cacciari:2011ma} to generate the standard anti-$k_T$ results.}

\begin{figure}[t]
\hspace{-1cm}
	\begin{tabular}{ccc}
&classical anti-$k_T$&$\alpha=10$\\[-1mm]
	\includegraphics[width=0.35\textwidth]{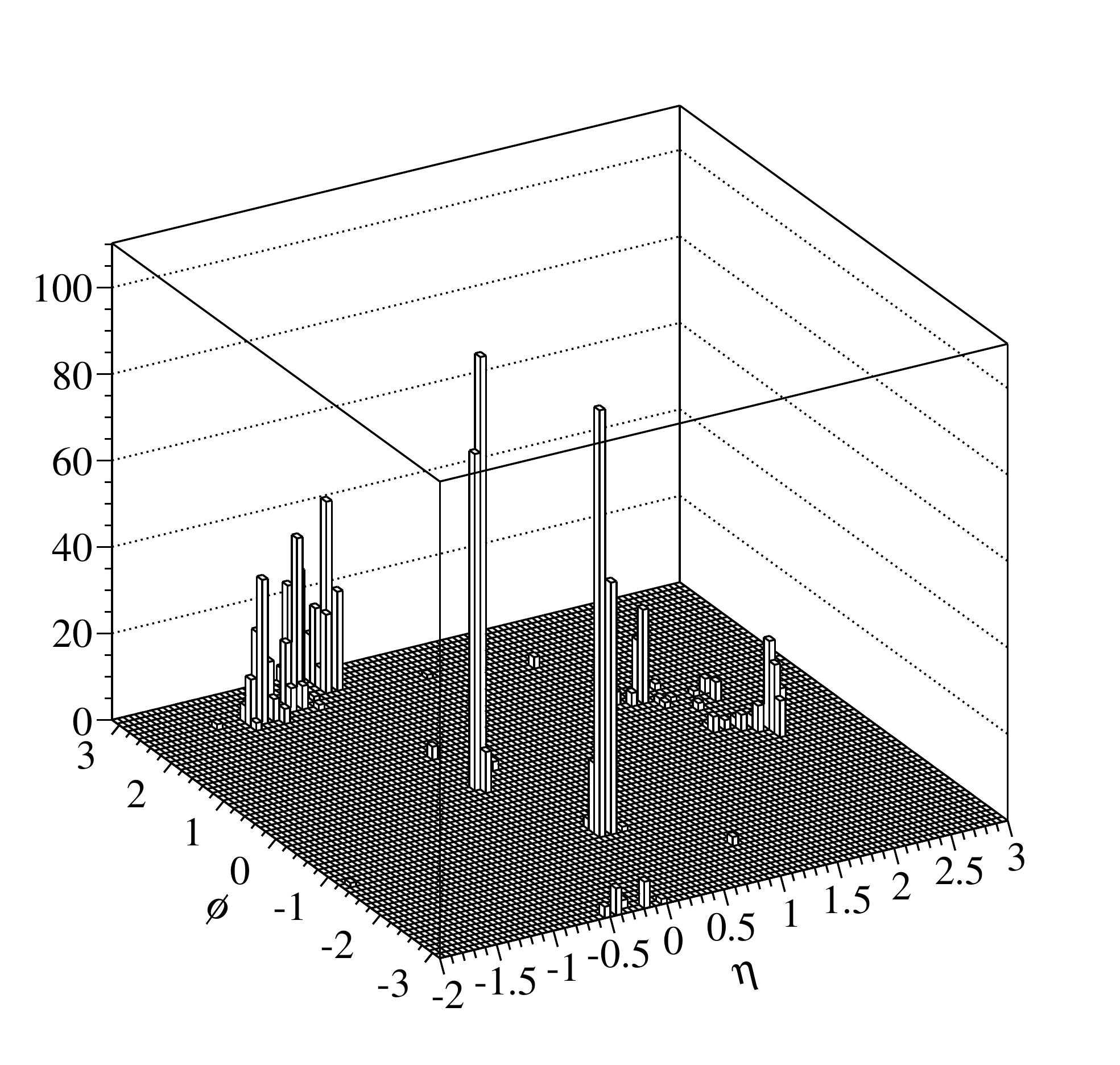} &
	\includegraphics[width=0.35\textwidth]{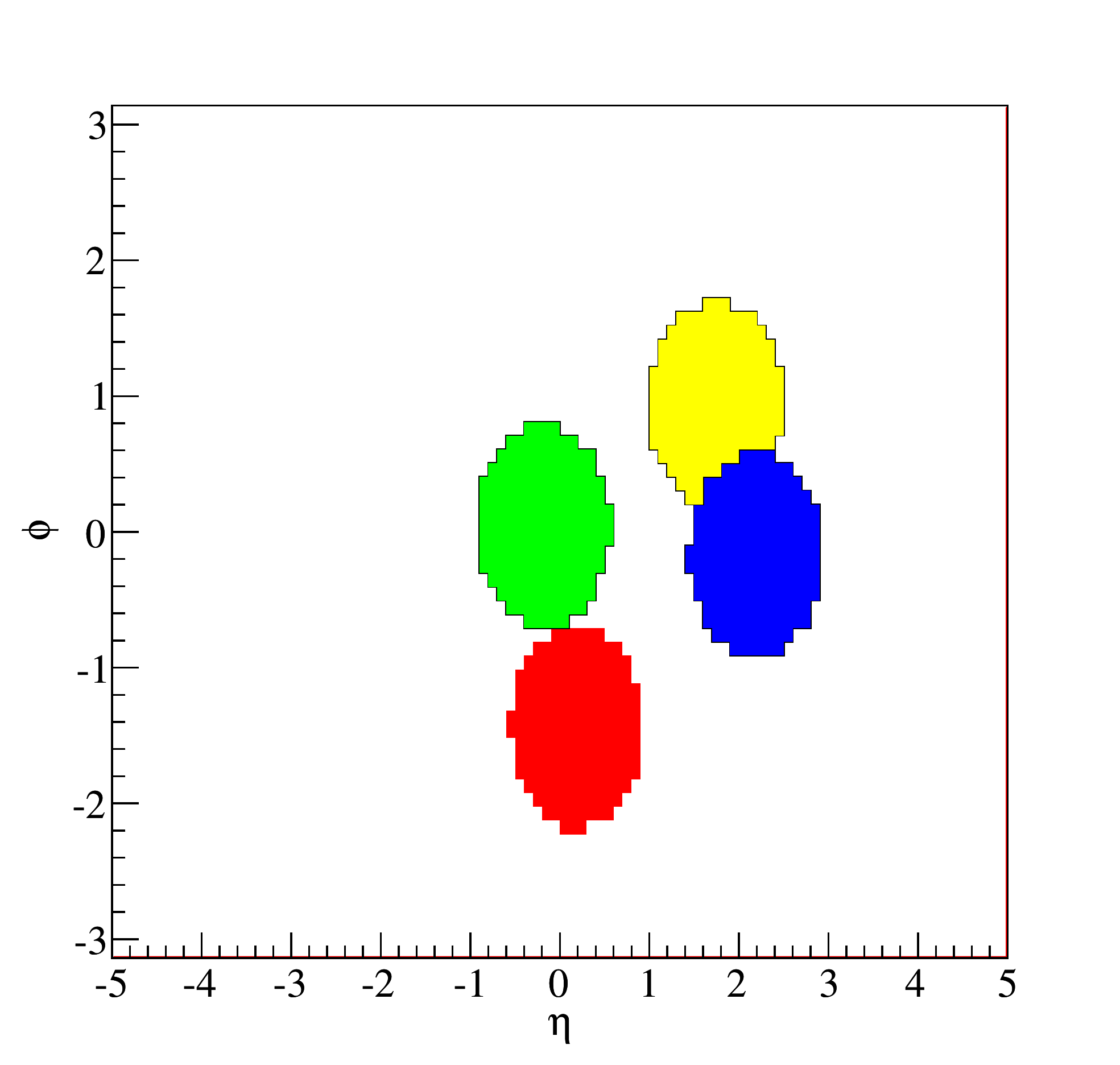} &
 	\includegraphics[width=0.35\textwidth]{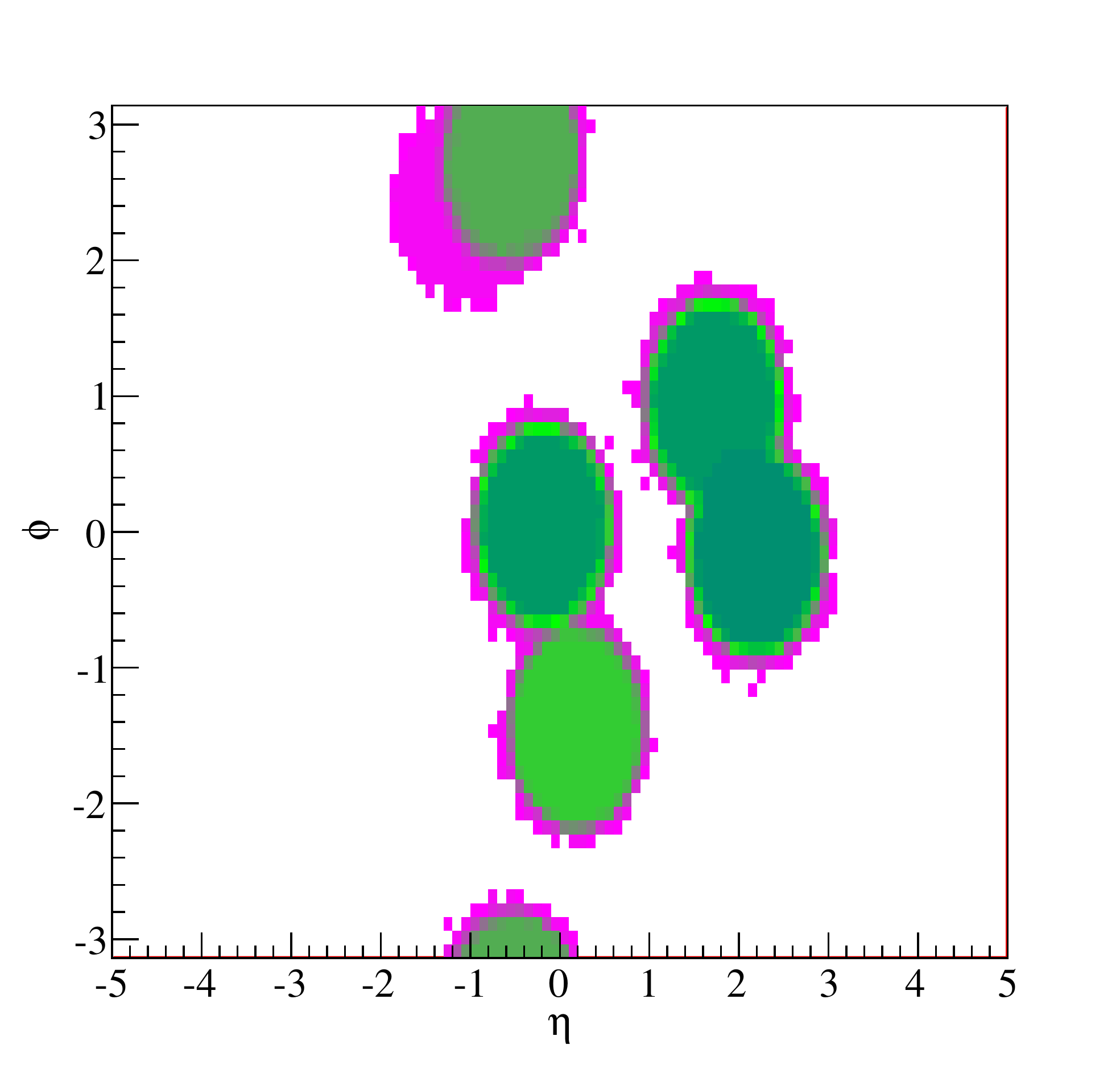}\\[2mm]
$\alpha=1.0$
&
$\alpha=0.1$
&
$\alpha=0.01$
\\[-1mm]
 	\includegraphics[width=0.35\textwidth]{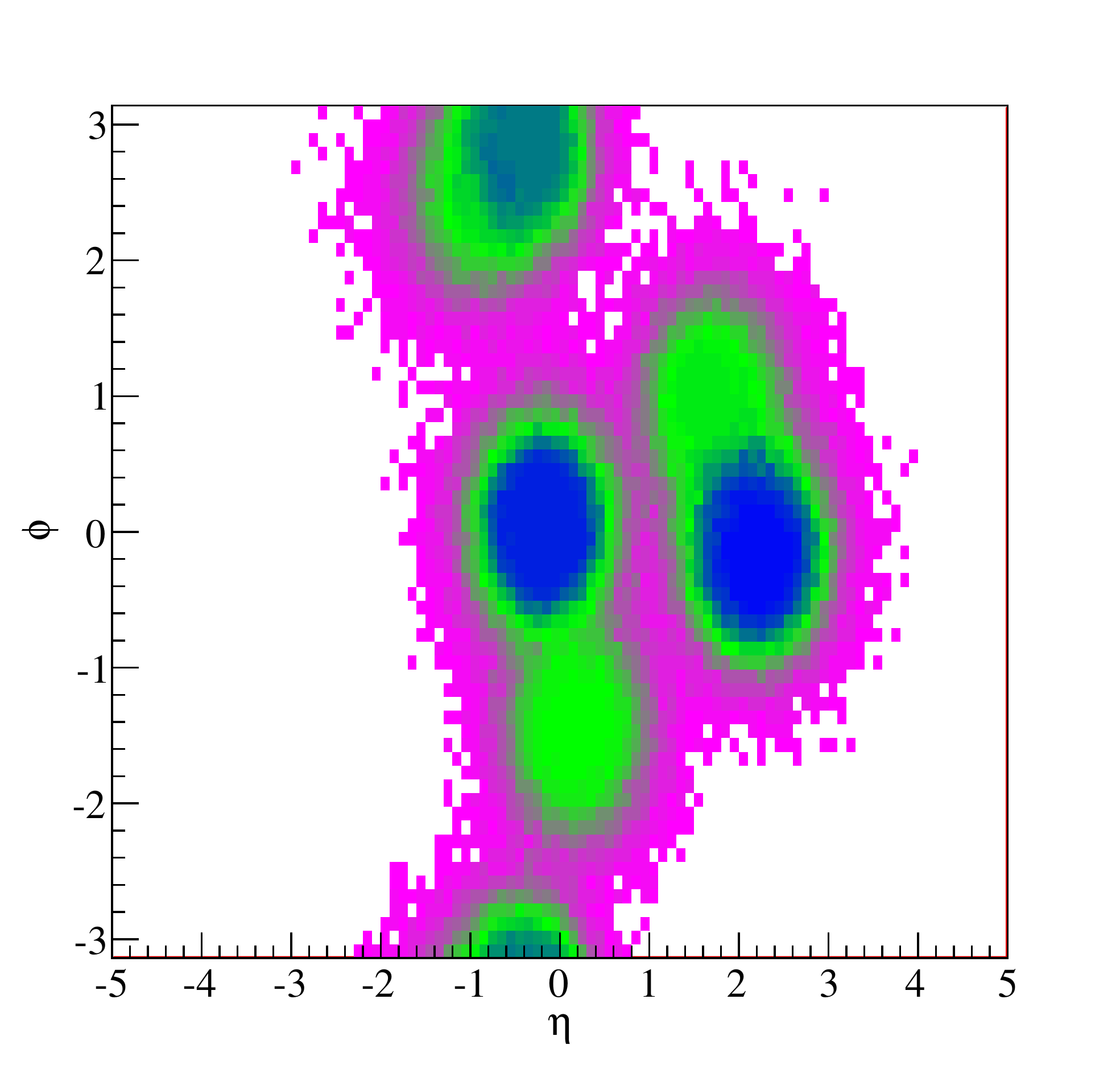}&
 	\includegraphics[width=0.35\textwidth]{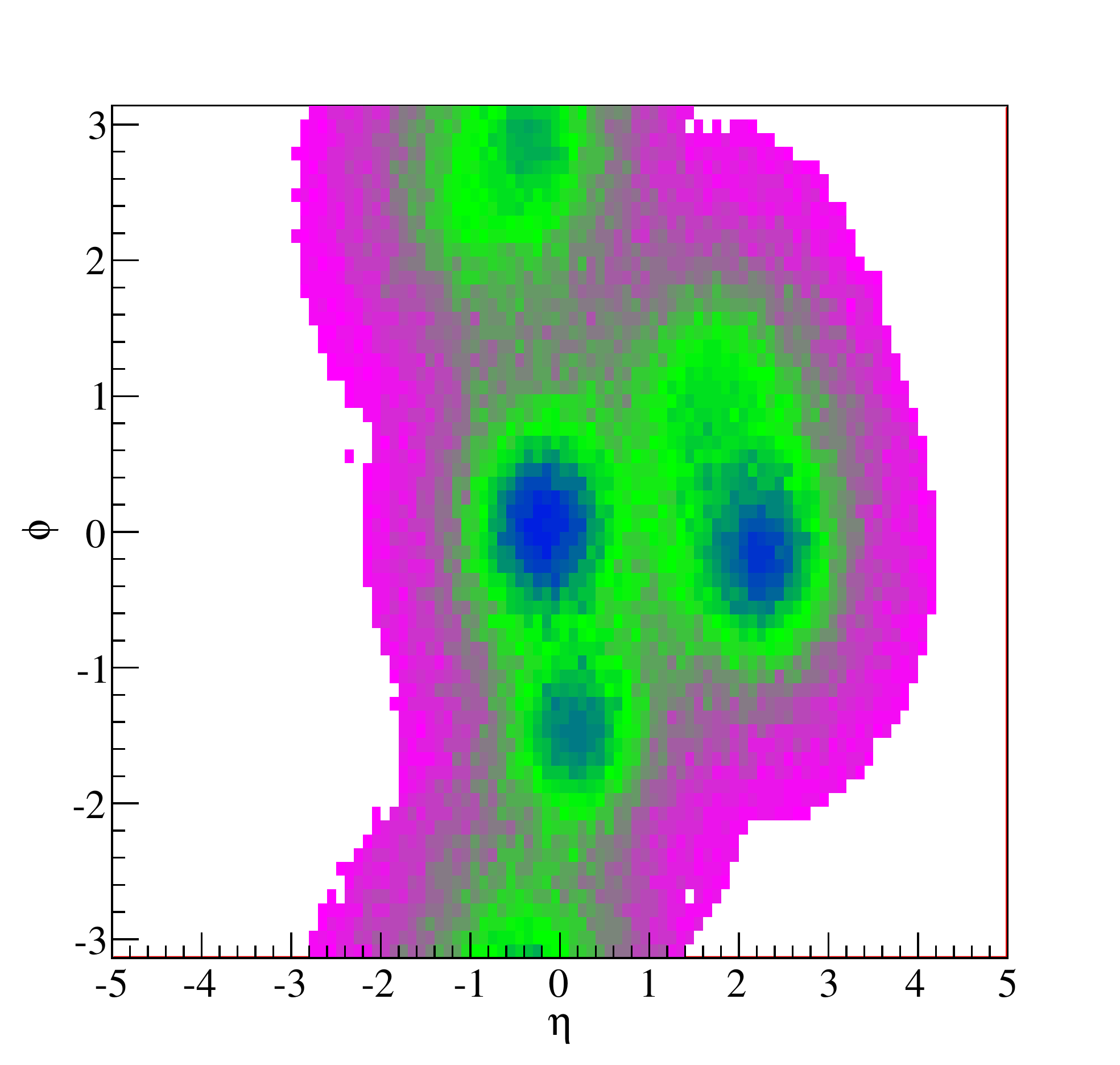}&
 	\includegraphics[width=0.35\textwidth]{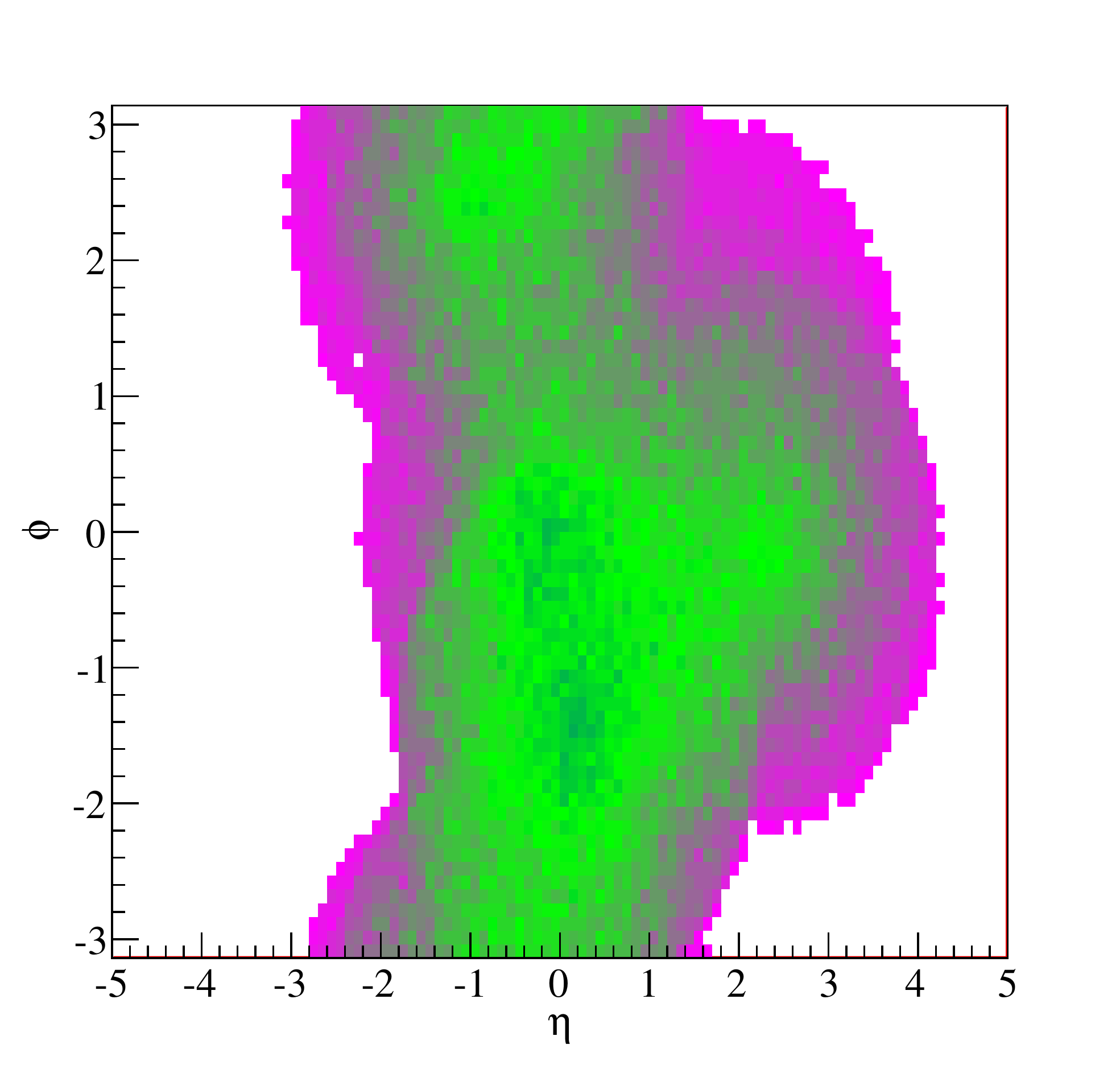}
  \end{tabular}
	\caption{The top-left panel shows the $\eta \times\phi$ plot of a simulated $pp \to \phi\phi \to gggg$ event at the LHC, with $m_\phi = 500\gev$.  The top middle panel shows
the jet areas associated with the four jets which best reconstruct the event using the
classical anti-$k_T$ algorithm (see Sec.~\ref{sec:fourjets}). The colors show the detector elements where zero-energy ghost particles would
get clustered into each jet. The remaining plots show the frequency with which a cell is clustered into one of the four jets which best reconstruct each event for different choices of $\alpha$.  
Blue squares indicate a cell is nearly always included amongst the four hardest jets, green squares indicate that the cell is included roughly half the time, while pink indicates a cell is only rarely included. 
% The values of $\alpha$ are, from top-left moving clockwise, $\alpha=10,1.0,0.1,0.01$.
The same event is shown in all plots.
 \label{fig:coloredplots} }
\end{figure}

\begin{figure}
	\centering
	\includegraphics[width=0.6\textwidth]{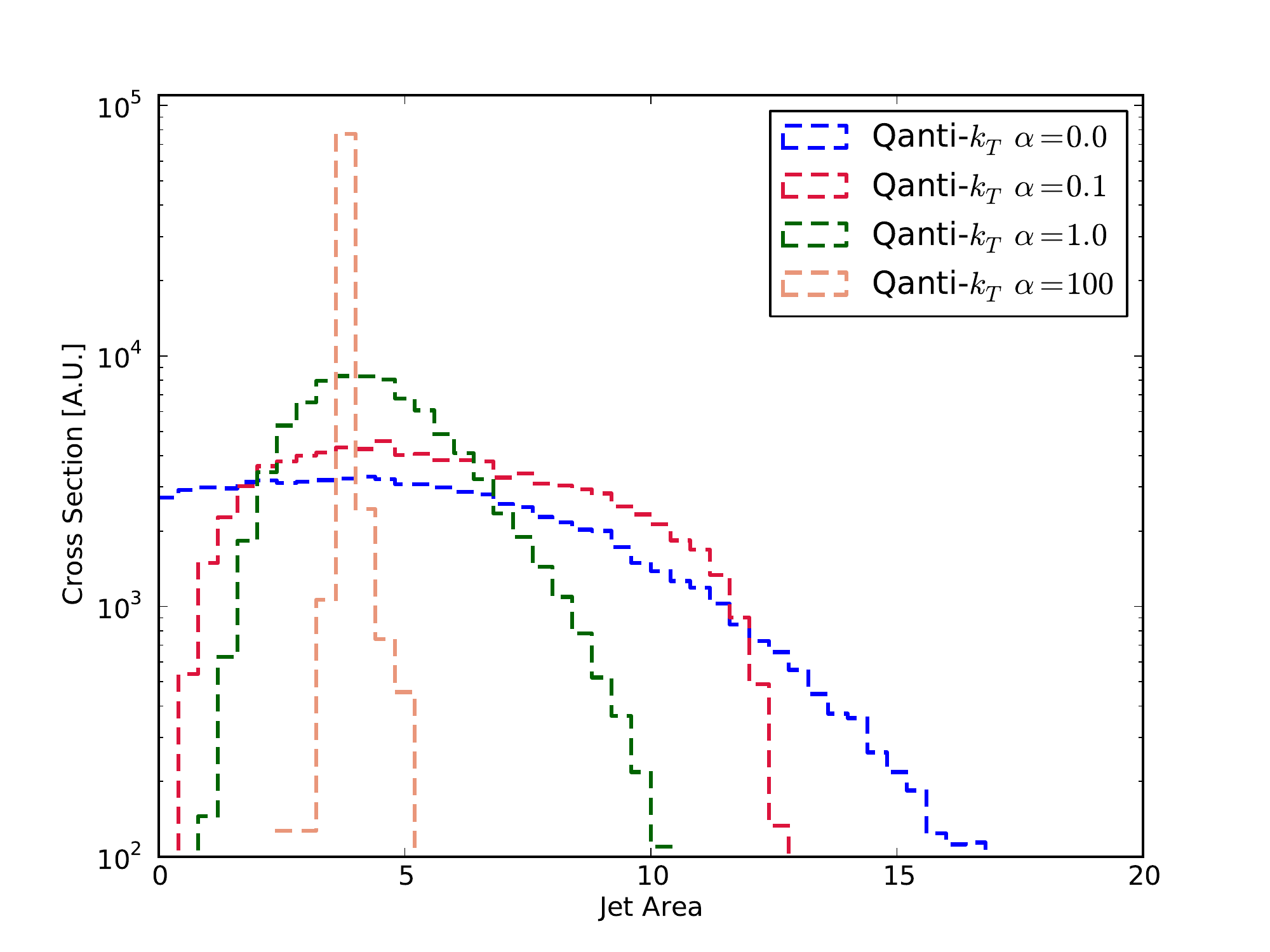}
	\caption{The jet area computed using Qanti-$k_T$ for various choices of the rigidity parameter $\alpha$. 
 Shown is the area of the hardest jet in $\phi\to gg$ dijet events with $m_\phi = 1$ TeV using $R=1.1$.\label{fig:jetarea}
 }
\end{figure}

To see how Qanti-$k_T$ handles overlapping jets, consider the four-jet event shown in Fig.~\ref{fig:coloredplots}.
This event is $pp \to \phi \phi \to gggg$ at the parton level, a process examined in Sec.~\ref{sec:fourjets}.
 In order to demonstrate that particles between jets can get clustered into different jets, we show what happens when ghost particles are added to the event.
 Ghost particles were introduced in~\cite{Cacciari:2008gn} as a way to characterize the area of a detector to which a jet is sensitive. 
Ghost particles are zero energy particles scattered throughout the acceptance region.
Since they have zero energy, they do not affect the location or 4-momentum of the final jets. 
The top-middle panel of~\ref{fig:coloredplots} shows the areas associated with the four jets which best reconstruct the event using classical anti-$k_T$ (see Sec.~\ref{sec:fourjets}).
This panel is similar to the bottom right panel of Fig. 1 of~\citep{Cacciari:2008gp}.

The remaining panels in Fig.~\ref{fig:coloredplots} show the frequency with which individual cells are clustered into the four jets which best reconstruct the event using Qjets for various $\alpha$. We see that for small values of $\alpha$ there is little well defined structure to the event, while for $\alpha=0.1$ we begin to see jetty areas of activity with amorphous borders.  Finally, for larger value of $\alpha$ we begin to resolve the 
standard anti-$k_T$ circular jet shapes.  Note in particular from the $\alpha=10$ panel that there are five jets relevant in this event
-- there is no clear choice between which four should be used in the reconstruction.  This is precisely the sort of ambiguity
which the multiple-interpretations approach can efficiently exploit.

One can be more quantitative about the area clustered into each jet using the jet area proposed in~\cite{Cacciari:2008gn}. In a classical
algorithm, this is the area of the detector clustered into a given jet. With Qjets, the area varies for each clustering. Thus the jet
area becomes a distribution. This distribution is shown in  Fig.~\ref{fig:jetarea}, averaged over many events for $R=1.1$. Jet area
using the classical anti-$k_T$ algorithm would give a $\delta$-function at $\text{area}=\pi R^2 = 3.8$. One can see this being approached
at large $\alpha$. For $\alpha =1.0, 0.1$ or $0$, the area is much broader. Thus, with Qanti-$k_T$, the jet area
can be either larger or smaller than what comes from using classical anti-$k_T$.

% \begin{figure}
% 	\centering
% 	\includegraphics[width=0.48\textwidth]{pt_760_p_0.pdf}
% 	\includegraphics[width=0.48\textwidth]{etaphi_classical_multicol760.pdf}
% 
% 	\caption{$\eta\times\phi$ plots of the calorimeter cells for a sample four-jet event. 
%  Left: here the area under each square is proportional to the $p_T$ of the cell. Right: the colored cells as those which compose the hardest four jets.  Contrast this plot with the same event clustered using Qanti-$k_T$ in Fig.~\ref{fig:coloredplots}.\label{fig:coloredplotsclassical}} 
% \end{figure}
% 
% \begin{figure}
% 	\centering
% 	\includegraphics[width=0.48\textwidth]{etaphi_760_p_5.pdf}
% 	\includegraphics[width=0.48\textwidth]{etaphi_760_p_6.pdf}
% 	\includegraphics[width=0.48\textwidth]{etaphi_760_p_7.pdf}
% 	\includegraphics[width=0.48\textwidth]{etaphi_760_p_8.pdf}
% 
% 	\caption{$\eta\times\phi$ plots of the calorimeter cells for the sample four-jet event shown in Fig.~\ref{fig:coloredplotsclassical}.  Here we show the frequency with which a cell is clustered into one of the four hardest jets in the event for different choices of $\alpha$.  Blue squares indicate a cell is nearly always included amongst the four hardest jets, green squares indicate that the cell is included roughly half the time, while pink indicates a cell is only rarely included. 
%  The values of $\alpha$ are, from top-left moving clockwise, $\alpha=0.01,0.1,1.0,$ and $10$. \label{fig:coloredplots} }
% \end{figure}

\section{Cut-weights}\label{sec:frac}
Once one generates $N$ clusterings of each event using Qanti-$k_T$, the clusterings can be used to improve the statistical significance in an analysis. In
the context of a search, combining multiple interpretations can be used to improve the  $S/\dB$ (the signal size divided by the characteristic background uncertainty) compared to a standard jet algorithm. Alternatively, the uncertainty on a mass, cross section, or branching
ratio measurement from a given sample can be reduced. In this paper, we focus on improving $S/\dB$.

\begin{figure}
\hspace{5mm}
      \begin{tabular}{cc}
%\hline
%classical anti-$k_T$& $\alpha=1$\\[-1mm]
	\includegraphics[width=0.45\textwidth]{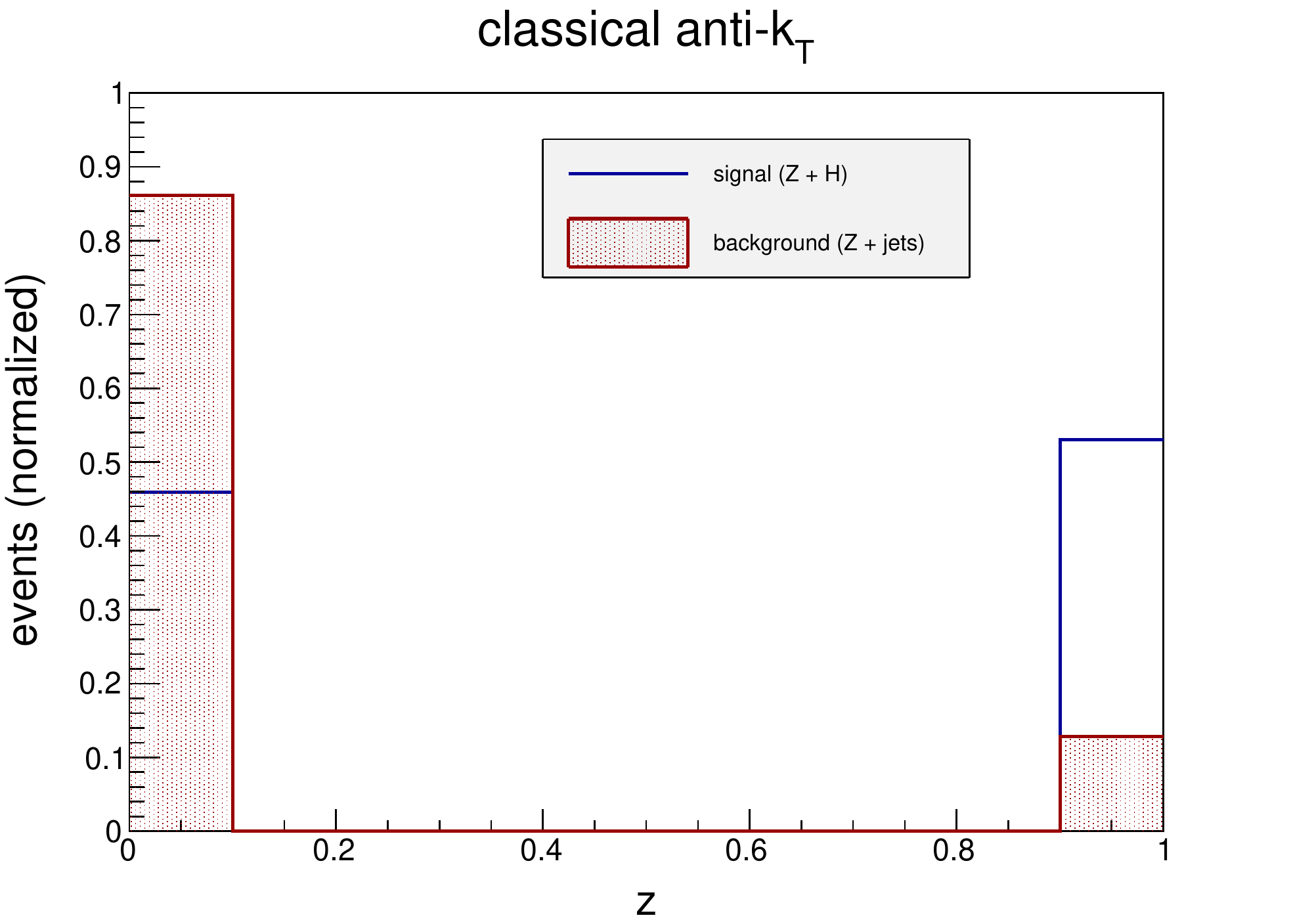}&
	\includegraphics[width=0.45\textwidth]{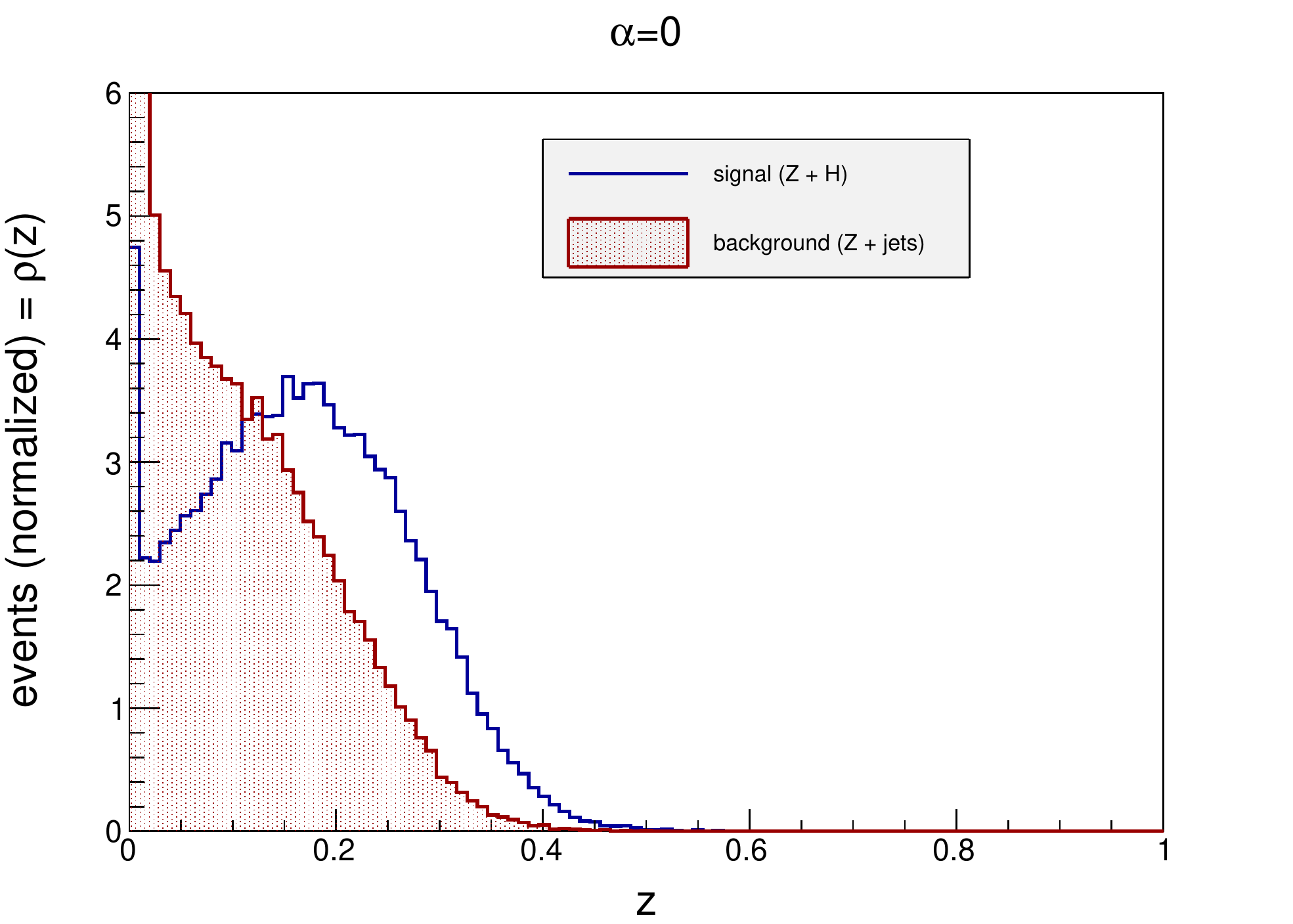}\\
%\hline
&\\
%[2mm]%$\alpha=0.1$&$\alpha=0.001$\\[-1mm]
	\includegraphics[width=0.45\textwidth]{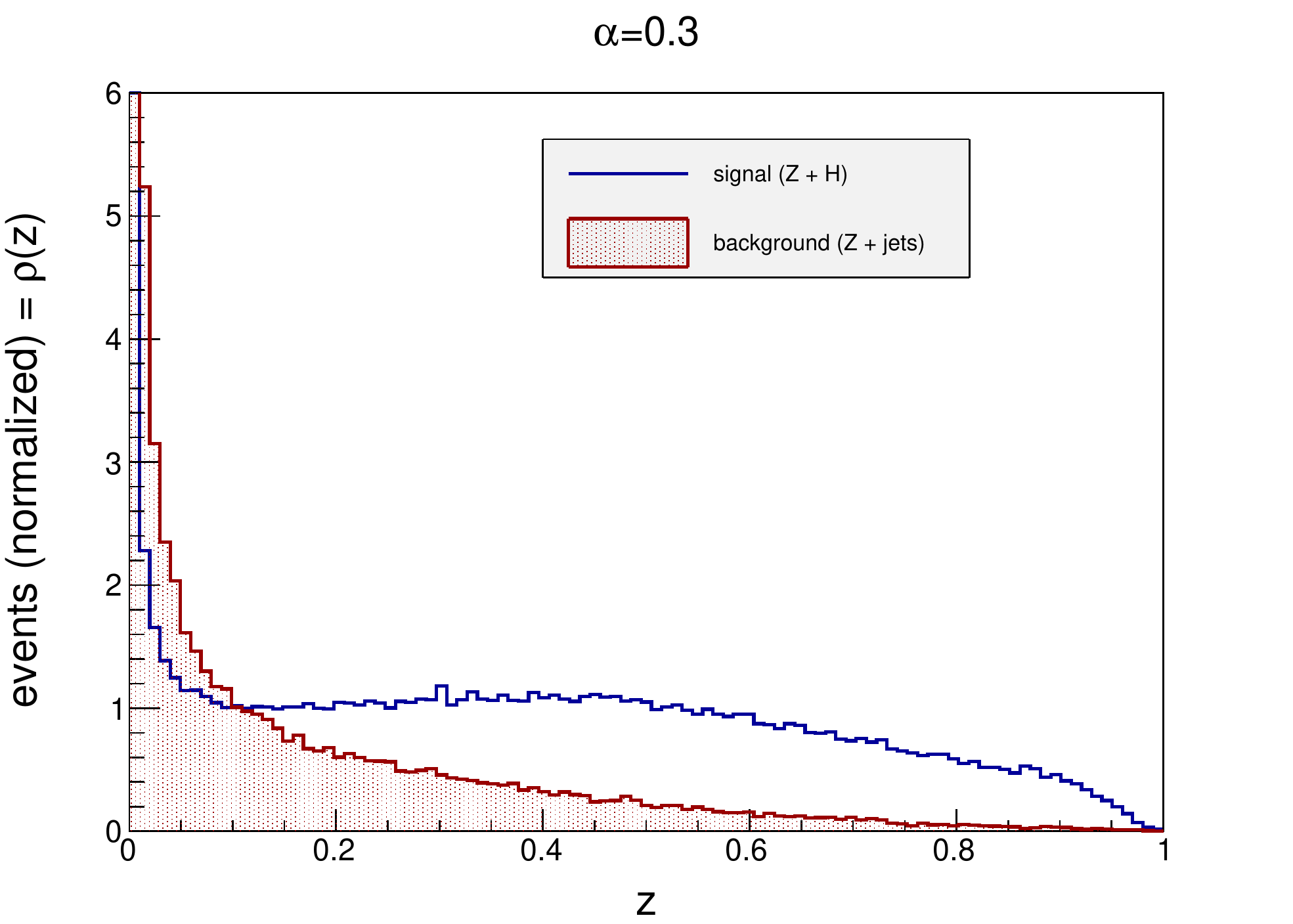}&
	\includegraphics[width=0.45\textwidth]{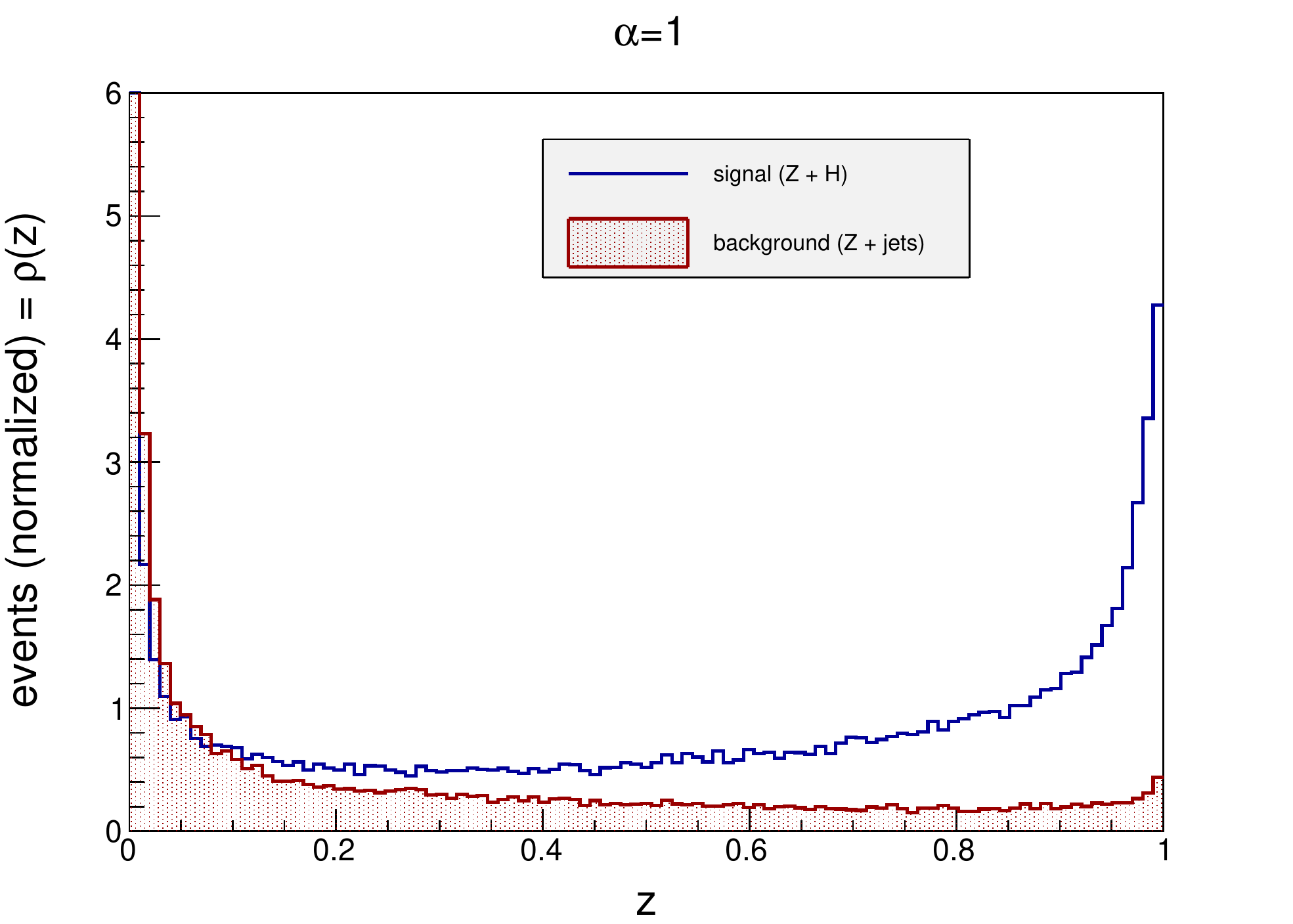}
\\
%\hline
    \end{tabular}
	\caption{
	$\z$ is defined as the fraction of interpretations of an event satisfying a set of cuts.
Shown is the distribution of $\z$ for signal ($H+Z$ events, hollow, blue) and background ($Z+b\bar{b}$ events, solid, red) for various $\alpha$. The cuts used to calculate $\z$ are 
$110~\gev< m_{JJ} < 140~\gev$ and $p_T > 25$ for each jet.
Top-left shows the classical case, where an event either satisfies the cuts $\z=0$ or it does not.
Distributions are normalized to area 1. These
normalized distributions are the functions $\f(\z)$ discussed in Sec.~\ref{sec:stats}.
 \label{fig:f1s} }
\end{figure}

\begin{figure}
\hspace{-12mm}
	\begin{tabular}{ccc}
	\includegraphics[width=0.35\textwidth]{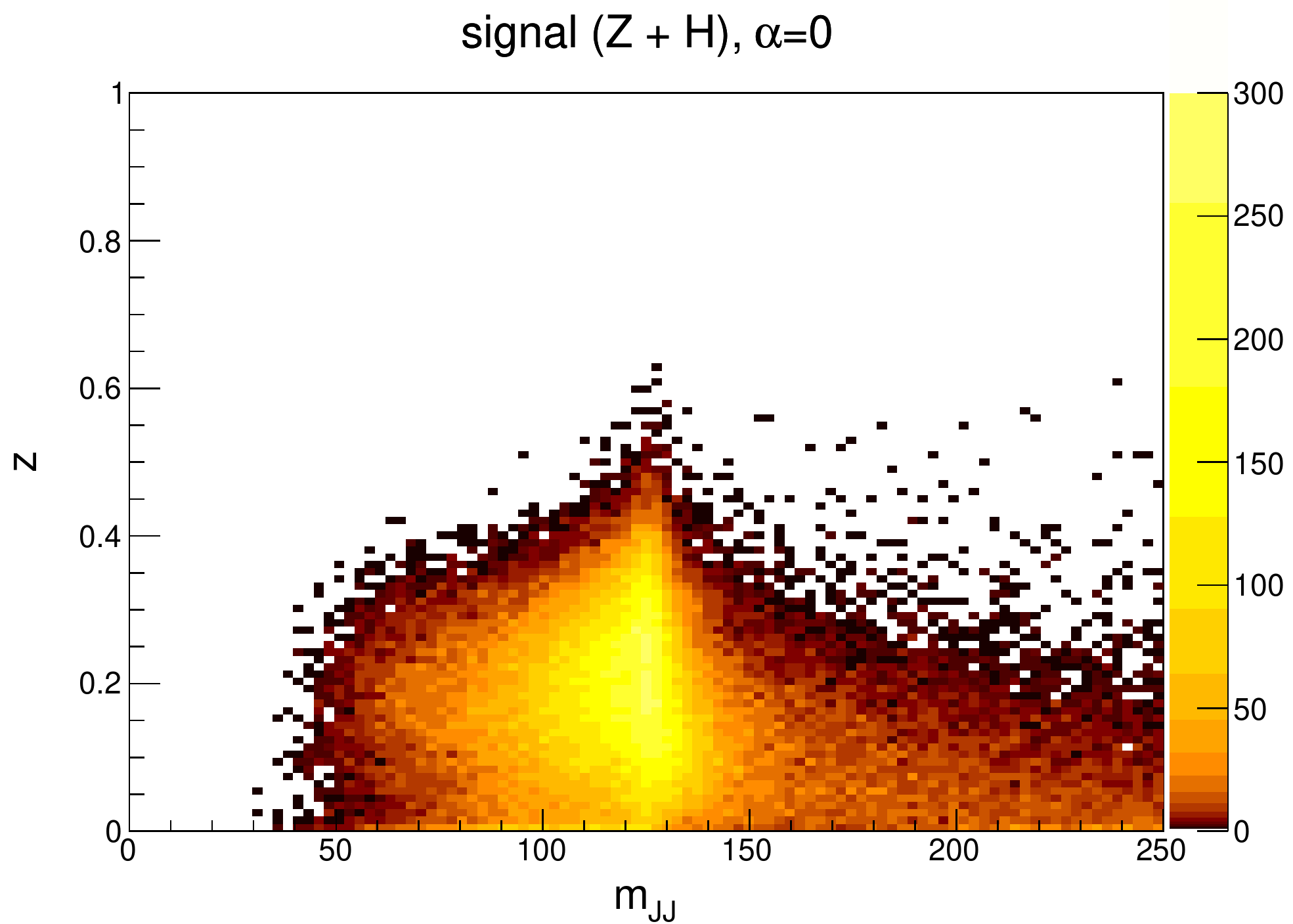}
&
	\includegraphics[width=0.35\textwidth]{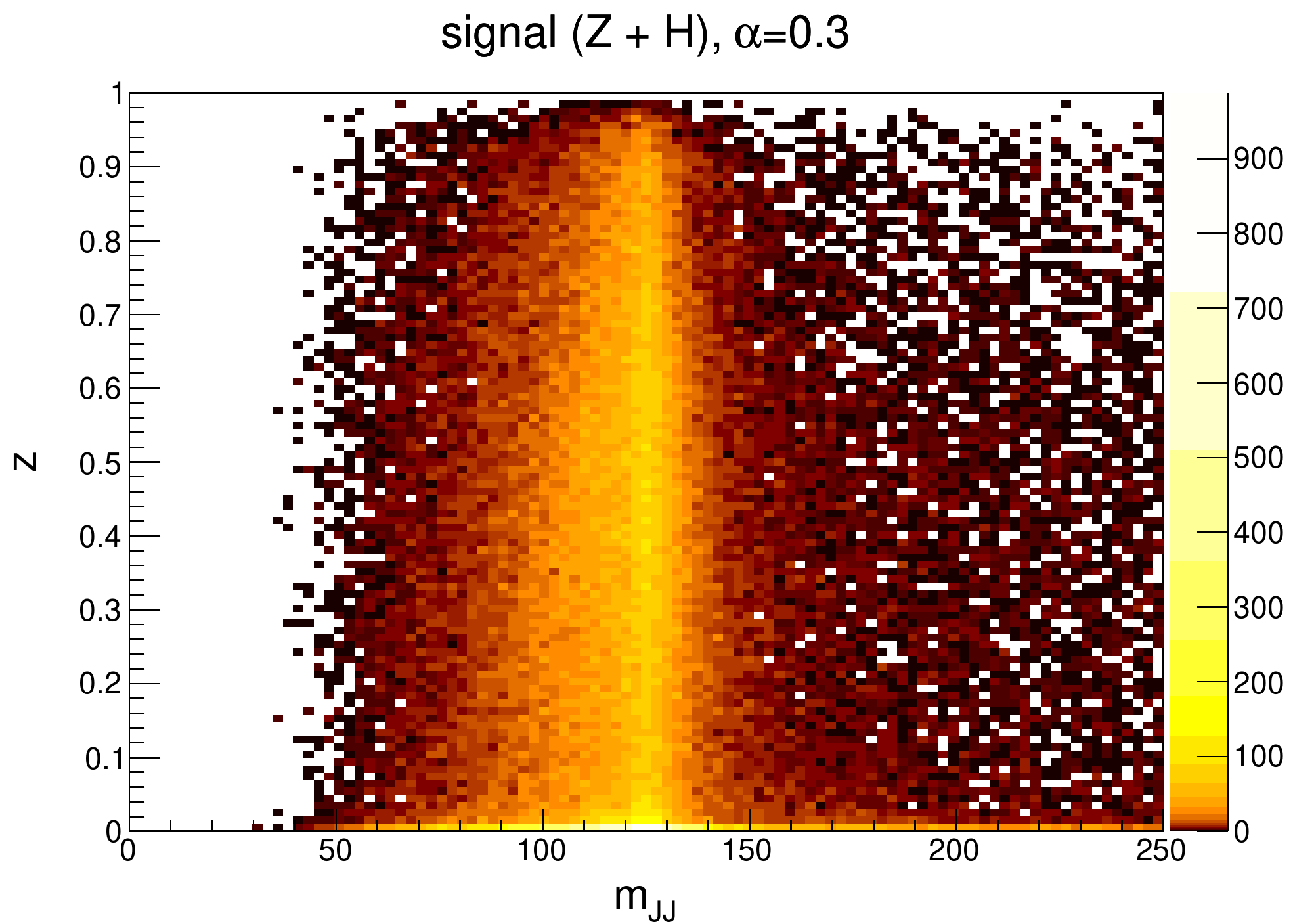}
&
	\includegraphics[width=0.35\textwidth]{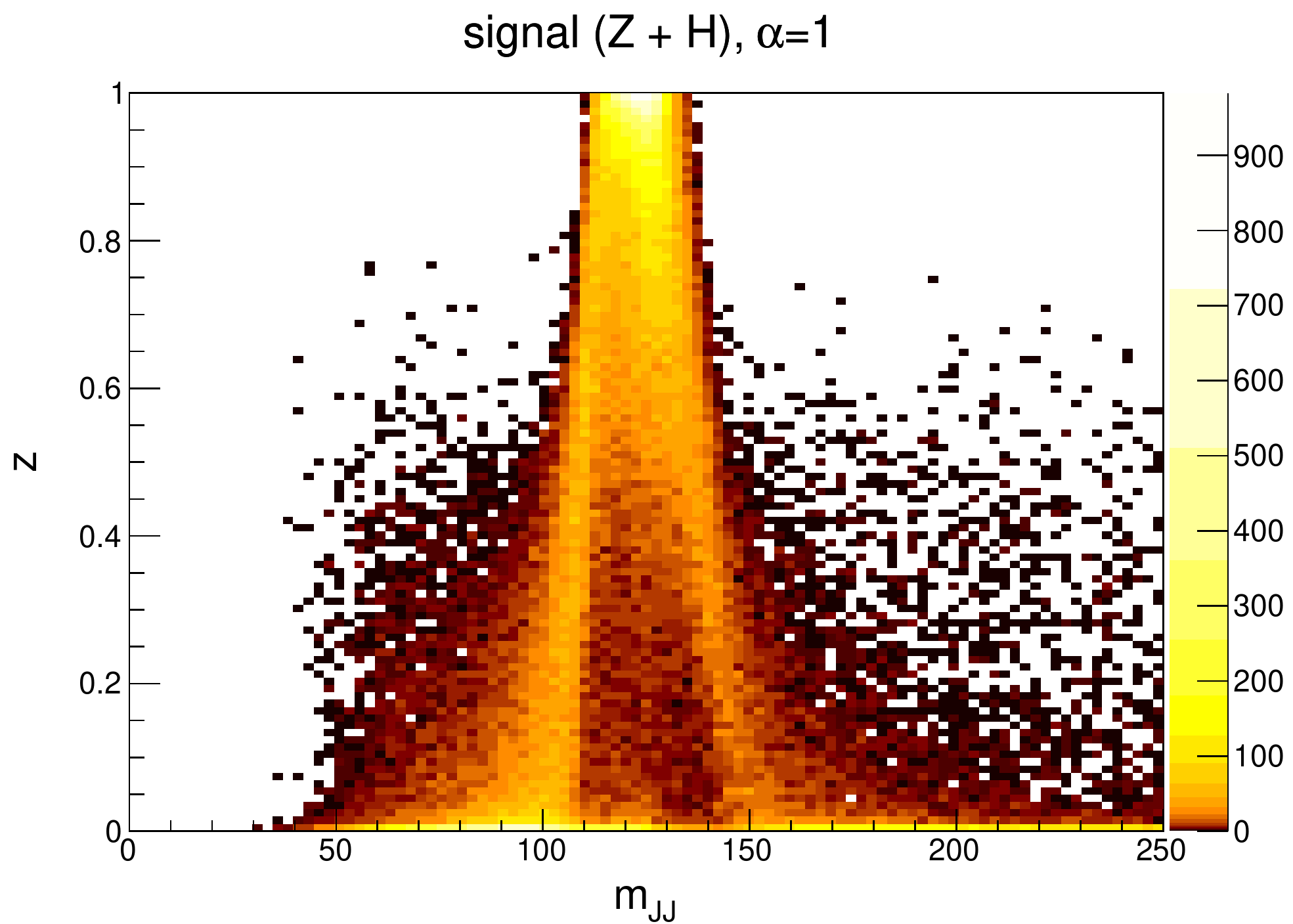}
\\
	\includegraphics[width=0.35\textwidth]{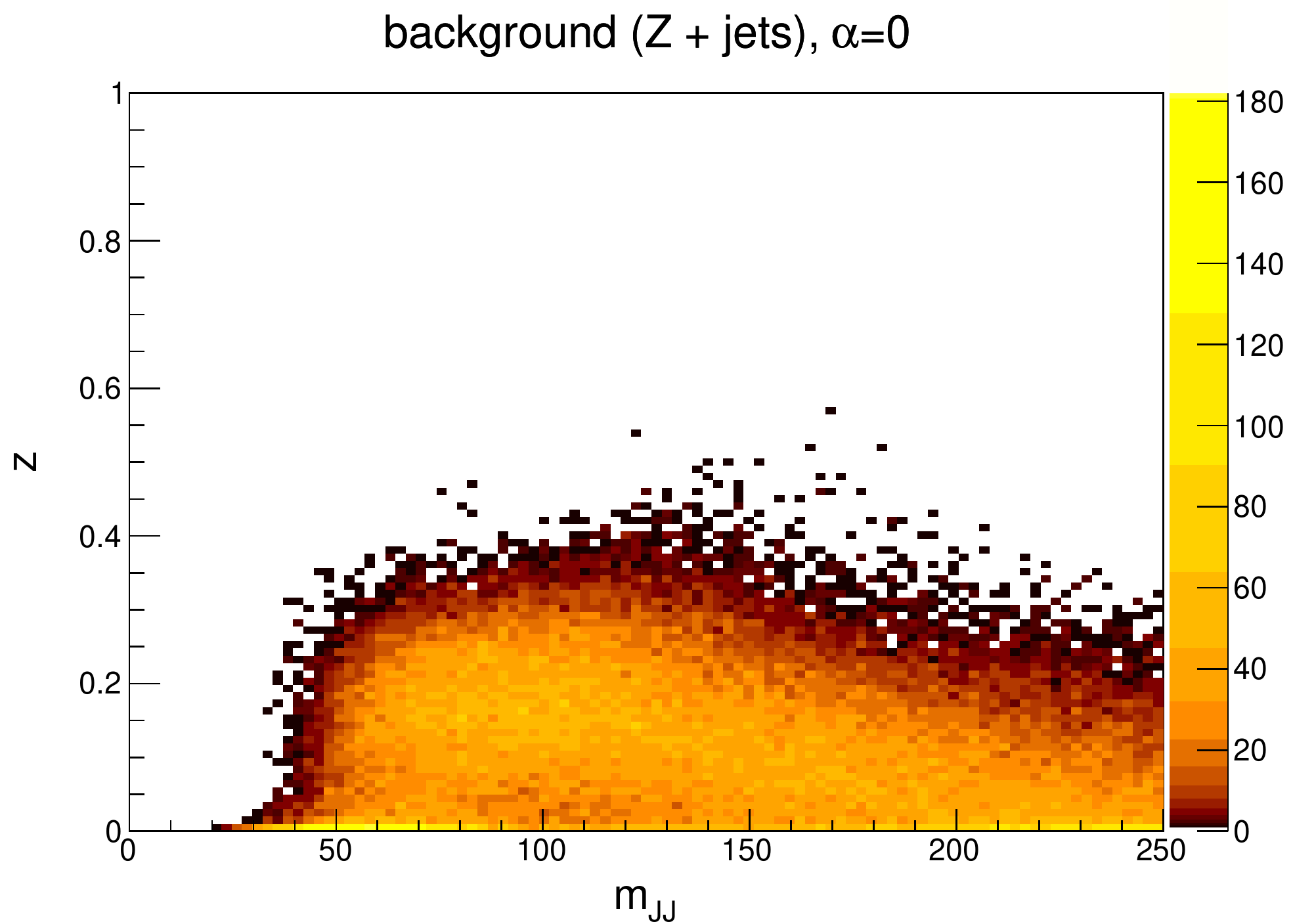}
&
	\includegraphics[width=0.35\textwidth]{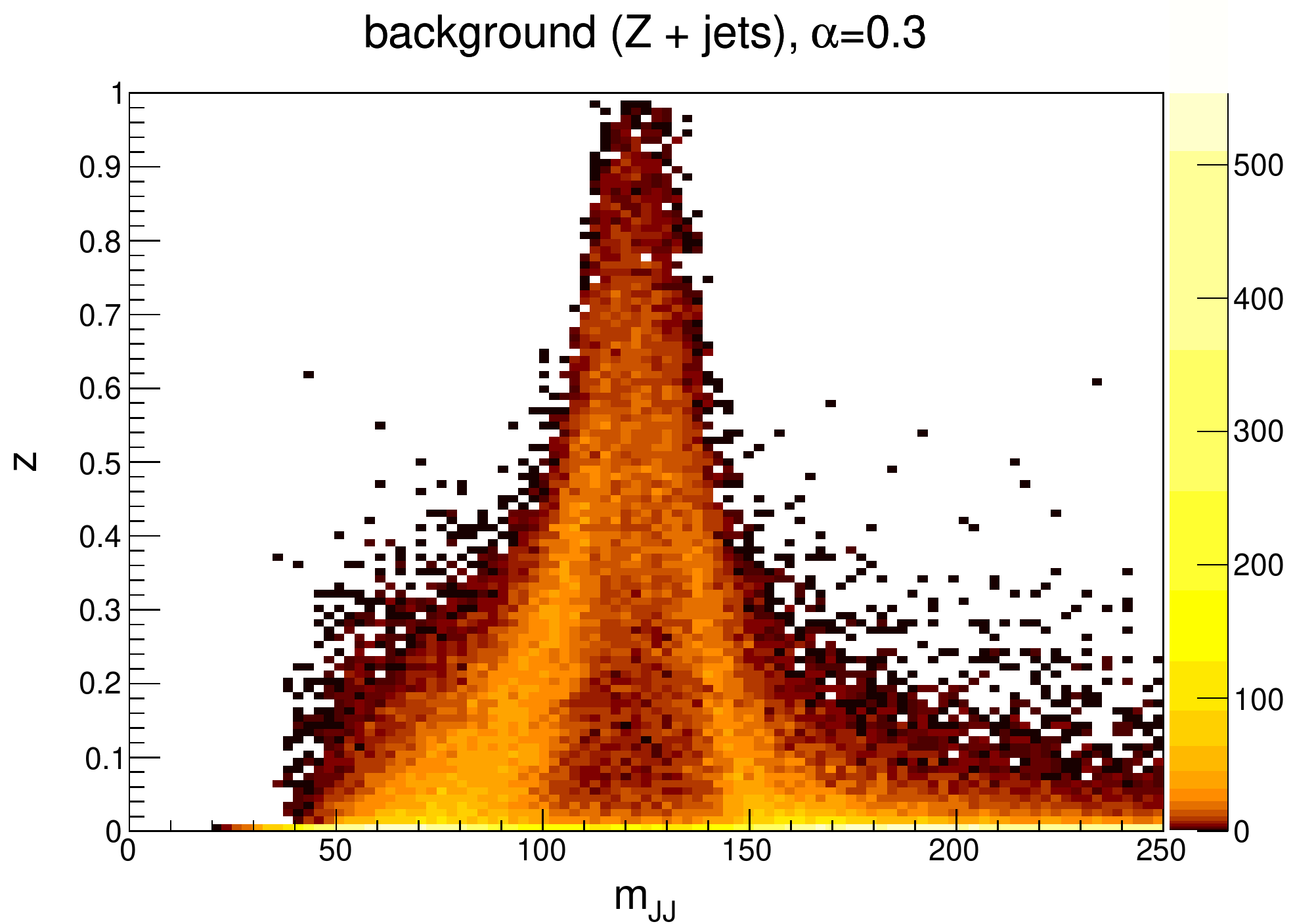}
&
	\includegraphics[width=0.35\textwidth]{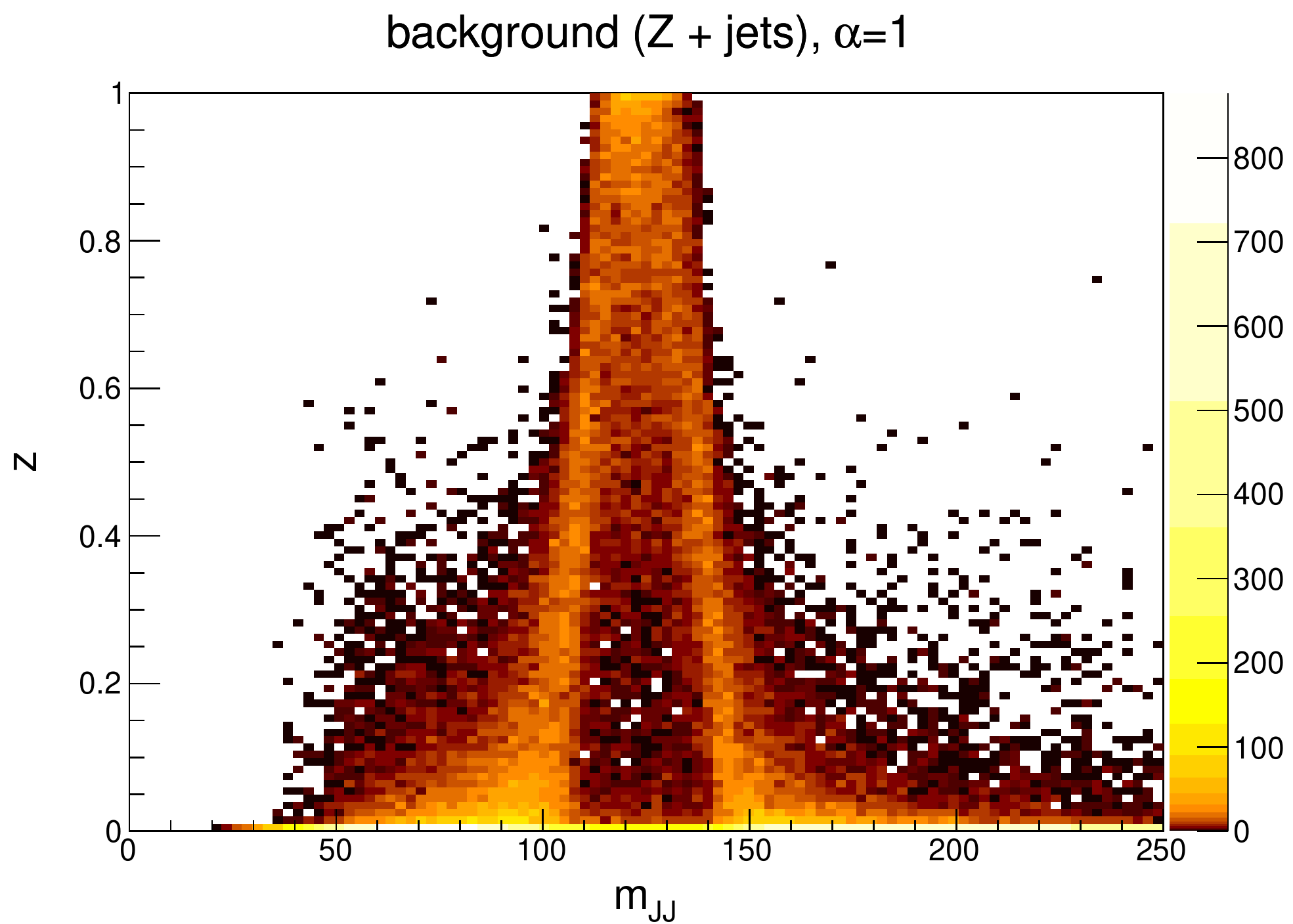}
\\

   	\includegraphics[width=0.35\textwidth]{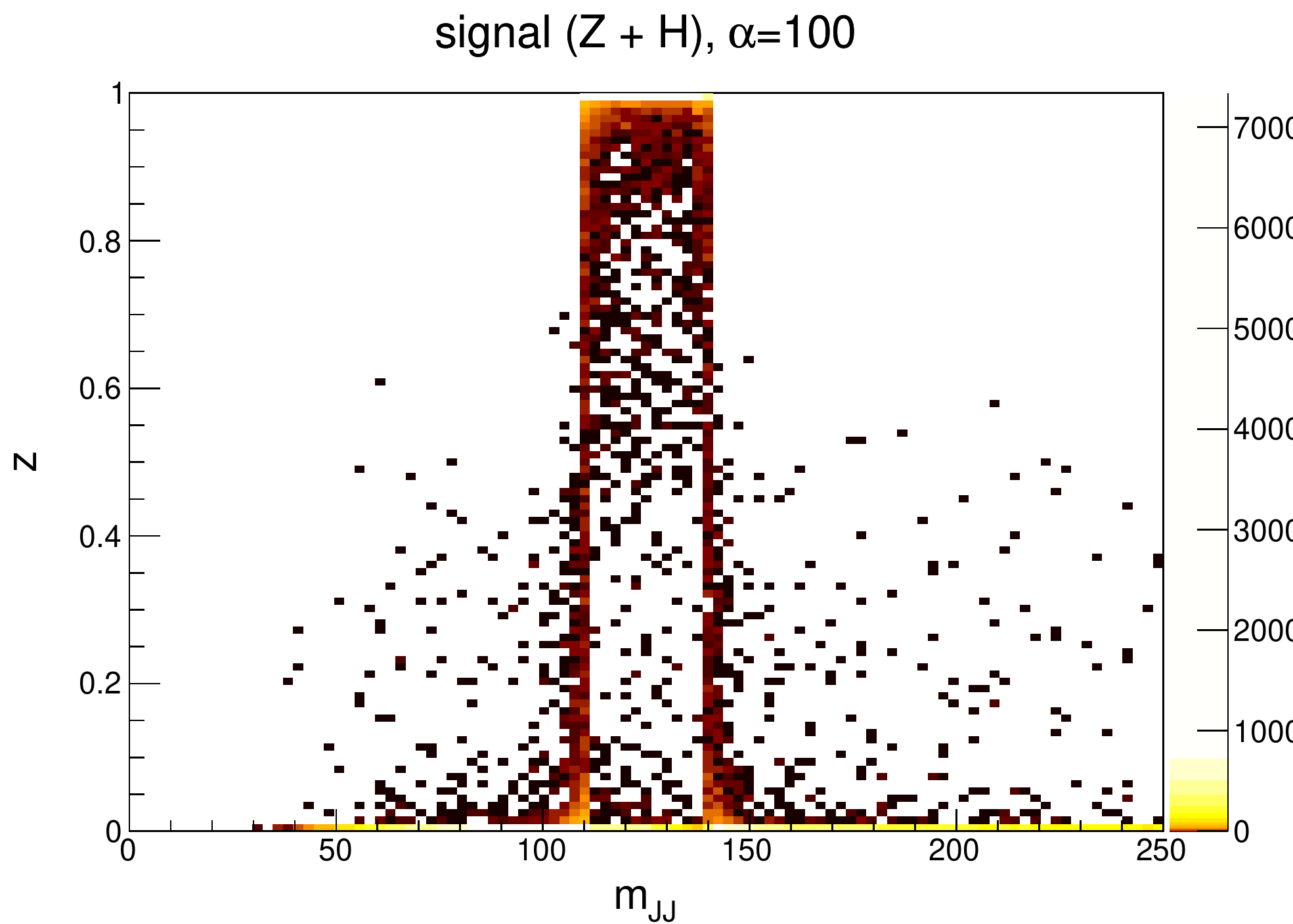}
  &
	\includegraphics[width=0.35\textwidth]{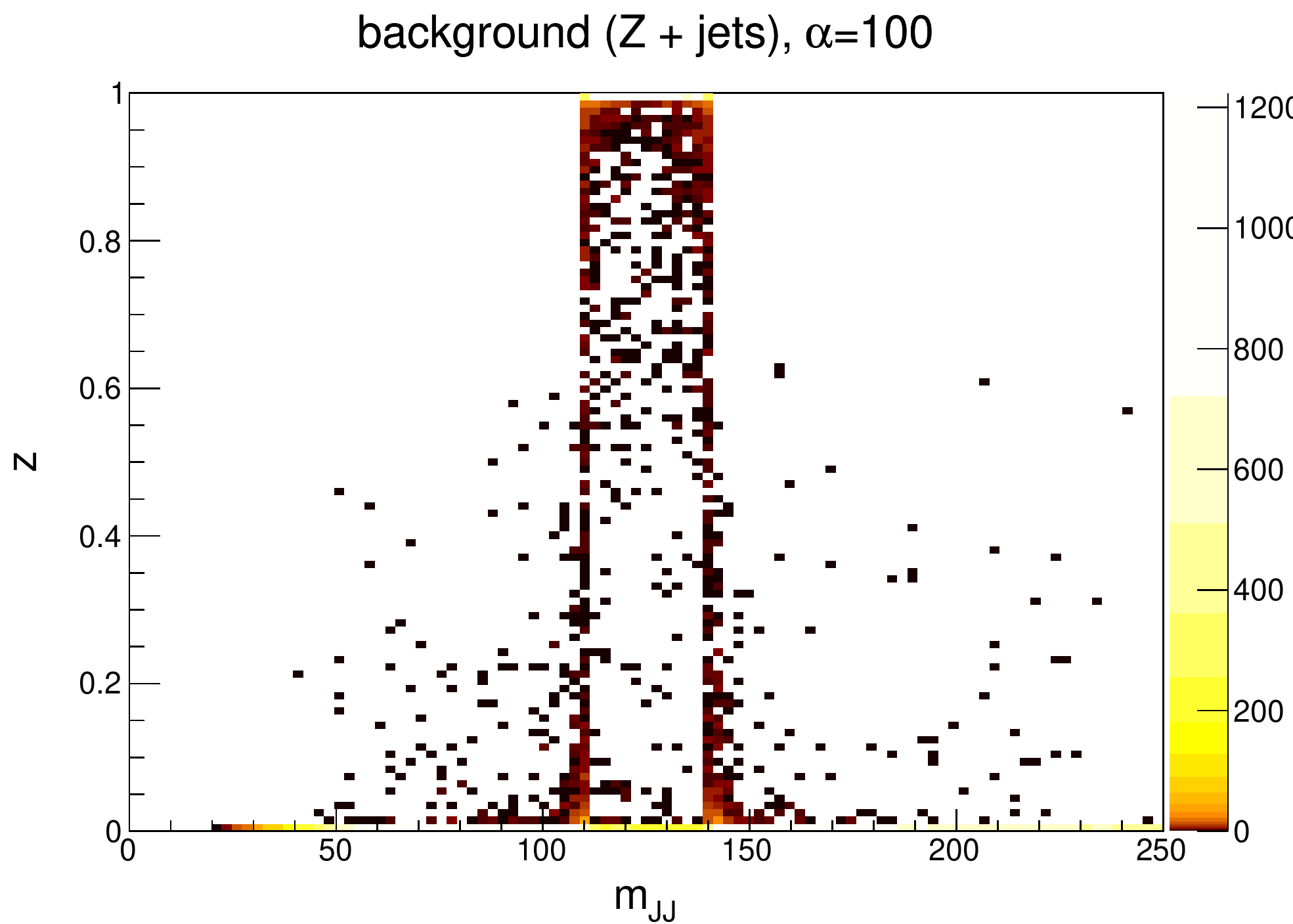}
  &
	\includegraphics[width=0.35\textwidth]{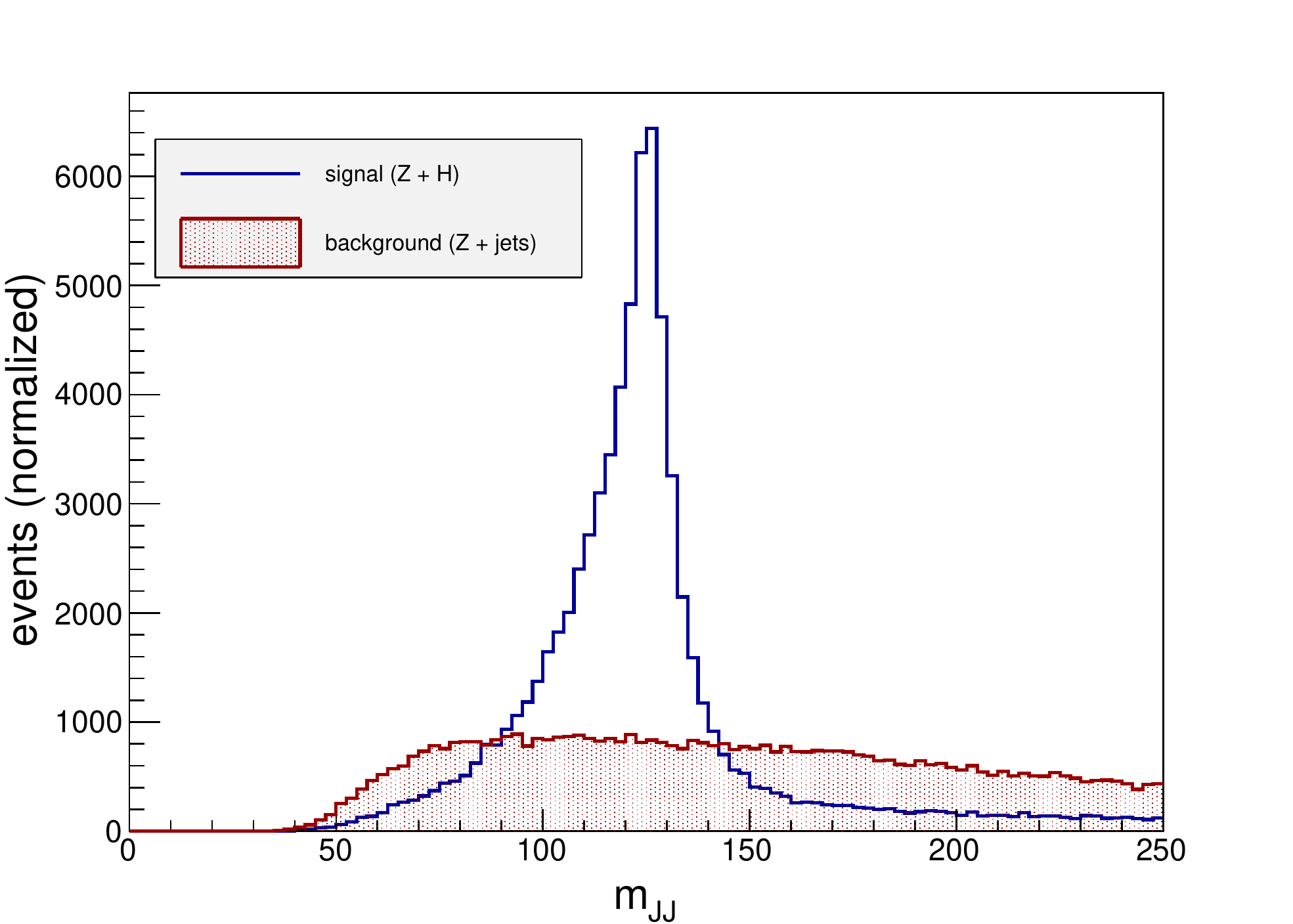}
  \end{tabular}
	\caption{$\z$ is the fraction of interpretations of an event which satisfy the cuts, as in Fig~\ref{fig:f1s}.
The 2D distribution of $\z$ as a function of the classical dijet mass $m_{JJ}$ is shown for some values of $\alpha$ for signal and background.
Every event gives a value of $m_{JJ}$ and a value of $0 \le \z \le 1$. Thus integrating over $\z$ reproduces the classical 
$m_{JJ}$ distribution, as shown in the bottom right. In the classical limit $(\alpha \to \infty)$, information from
multiple interpretations is inaccessible.
\label{fig:xivsmjj}} 
\end{figure}

Suppose one decides on a set of cuts which optimally distinguish signal from background for a particular classical analysis. For example, in searching for $H+Z$ events with $H\to b\bar{b}$ one might like to cut on the invariant mass of the $b\bar{b}$ pair. Whatever the cuts are, classically an event either passes those cuts or does not pass them. With Qanti-$k_T$, a fraction $\z$ which we call the {\bf cut-weight}, of the events can pass the cuts.

To get a feel for what the cut-weight distributions look like, we show signal and background distributions of $\z$ in Fig.~\ref{fig:f1s}. 
Here, signal is $H+Z$ events with $H\to b\bar{b}$ and background is $Z+b\bar{b}$ events. We demand that $p_T^Z > 120\gev$ for all
events, imagining $Z$ decays to neutrinos and this is a missing transverse energy cut. The cuts by which $\z$ is determined are
that the two hardest jets should have $p_T> 25\gev$ and that the dijet invariant mass of the two hardest jets is in the window
$110\gev < m_{JJ} < 140\gev$.

In the classical limit ($\alpha=\infty$), we see that $\z = 0$ or $\z = 1$ only. That is, an event either satisfies the cuts or
it does not. For smaller $\alpha$, say $\alpha=1$, there is a substantial fraction of the events for which only
some of the clusterings satisfy the cuts. Note that more signal events pass the cuts than background events. 
For $\alpha=0$ where the clustering is random, no more than half of the interpretations pass the cuts. 

As another way to visualize the value added by cut-weights, we show in Fig.~\ref{fig:xivsmjj} how $\z$ changes for events with
a given classical dijet invariant mass. In a classical analysis, one can look at the distribution of $m_{JJ}$ for signal and
background and put a cut to optimize significance. Such a cut corresponds to a vertical band in these plots. Because the distribution
of $\z$ is different for signal and background events with the same value of $m_{JJ}$ it will help to incorporate $z$ into the 
analysis. Applying a 2 dimensional cut on $m_{JJ}$ and $\z$ provides around a 6\% improvement (for $\alpha=0.1$) in $S/\sqrt{B}$ 
over a classical (vertical) cuts
on the same event. Combining $m_{JJ}$ with $\z$ using both $\alpha=1$ and $\alpha=0.1$ using boosted decision trees 
gives around a 7\% improvement. However, cutting on $\z$ is not ideal since one is still throwing out events instead of
weighting the less signal-like events less. We discuss next how to compute the significance using weighted events.

\section{Statistics} \label{sec:stats}
The fraction $z$ of events passing a set of cuts provides a weight for each event based on how many interpretations of that event resemble signal according to some measure.
Thus it is natural to use these weights directly in the calculation of the significance. In this section we discuss how this can be done.
The procedure we describe here was used in~\cite{Ellis:2012sn} and is discussed in more detail in~\cite{Qstats}.

If one knew what the signal and background distributions should look like exactly, the optimal significance would be achieved by using something like a likelihood test. In practice, we never know {\emph{exactly}} what signal and background should look like. Thus
using likelihood ratios can be prone to picking up on pathological regions of simulations.
Moreover, it can be extremely challenging to calculate the systematic uncertainty on likelihood-based significance estimates. Cuts provide a compromise where
the simulation does not have to be perfect and the systematic uncertainty can be estimated more reliably. 
Multiple interpretations through the Qjets approach provides a method for combining some of the advantages of both the cut-based and  likelihood-based approaches. By using the fraction of interpretations in a window as a cut,
 one knows explicitly what regions of phase space are contributing (as in a cut-based approach). However, since events that are more signal-like contribute more,
the significance of an excess will be greater for a given luminosity than using a cut-based approach alone.

\subsection{Significance}
To quantity the improvement from our procedure we adopt as a measure the 
excess number of events measured $S$  divided by the expected fluctuations
in the background $\dB$. That is,
\begin{equation}
\text{significance} = \frac{S}{\dB} =
 \frac{N_{\text{observed}} - N^\text{bkg}_{\text{expected}}}
 {\delta N^\text{bkg}_{\text{expected}}}
\end{equation}
%This can be approximated in perturbation theory or in simulation. 
For example, suppose we see $S=100$ excess events in some
channel which  a Higgs boson could contribute to. If the background was only expected to fluctuate
by $\dB = 20$ events, then the significance is $S/\dB=5$, which conventionally
characterizes a discovery. That is, in order to replicate the observed number of events, the background
without signal would have had to have fluctuated by 5 times more than $\dB$.
To calculate the significance with data, one needs to know the mean and variance of $N^\text{bkg}_{\text{expected}}$. 

A key feature of Qjets is that events are not characterized as signal or background (e.g. by
passing some cuts or not passing them). Rather they are assigned a weight $\z$ between 0 and 1 based on how many interpretations of the event are signal-like (according to some measure). Thus the measured number of signal events $S$ no longer
has to be an integer. Moreover, the fluctuations in the background are now the fluctuations in
a non-integer number. 

The practical procedure we propose is very simple: count the number of events passing a set of cuts weighted by $\z$. That is,  define $N_\text{observed}$
as the sum over $\z$ for each event (rather than counting the number of events with $\z=1$). In order to decide if this number is consistent with a background-only hypothesis, one needs to know the expected fluctuations in this weighted number. We now describe how the expected size of fluctuations can be easily computed.
% 
% Before going into the details of the statistics, it is worth emphasizing that the simulation of the signal process is only important in designing the cuts to enhance the expected excess. In deciding if an excess exists in data over background, one only needs an accurate calculation of the distribution off $B=N^\text{bkg}_{\text{expected}}$.
% 
We first review how the expected value and variance of $B$ are computed in a classical
analysis and then describe how cut-weights can improve significance.

\subsection{Classical cut-based significance} 
Suppose we are looking for a particular signal (like a Higgs boson) in a classical analysis
and we design a set of cuts to optimize the discovery potential. Once the cuts are set, we can focus on the background expectation and fluctuations, since these will determine the significance of an observed excess. Let us say with a given
luminosity that we expect $N$  background events of a particular type to be produced. 
Let us say a fraction $\epsilon$ of these background events are expected to pass a set of cuts.
We call $\epsilon$ the {\bf reconstruction efficiency}.
Thus, in the absence of signal, we expect $N\epsilon$ events. We would next like
to know what the expected variance is around this mean. 
There are two contributions to the fluctuations about the mean: from the inherent quantum mechanical Poisson process which produces the events
in the first place, and from the fact that any individual event has some probability of satisfying
our cuts.

The production rate is governed by a Poisson distribution. If we expect $N$ events, 
the probability of producing $n$ events instead is
\begin{equation}
P(n|N) = \frac{e^{-n}n^N}{N!}
\end{equation}
This Poisson distribution has mean $N$ and standard deviation $\sigma = \sqrt{N}$. The variance is $\sigma^2=N$.
% (recall that the variance is the square of the standard deviation).
%In the next phase, we use a jet algorithm to reconstruct the event. For concreteness we consider an example where a decaying scalar $\phi$ of mass 500 GeV is reconstructed by a jet algorithm. We require a successfully reconstructed scalar to have a mass that falls in the window 450 - 550 GeV.

Now consider the reconstruction efficiency. Say our background events pass our cuts a
fraction $\epsilon$ of the time. For example, for the samples shown in Fig.~\ref{fig:f1s},
we can see from the top-left panel (the classical case) that $\epsilon_B=0.12$ for background
and $\epsilon_S=0.55$ for signal.
Suppose there is only signal. 
If $n$ signal events are produced, what is the probability of finding $a$ events passing our cuts?
%
%
%To model variations due to events not passing our cuts.
%
% A classical jet algorithm such as anti-$k_T$ is able to do this with some efficiency $\epsilon$. For instance if $\epsilon = 0.5$, the algorithm will correctly reconstruct the scalar half of the time. 
% 
It is not hard to see the this probability is given by a weighted binomial distribution:
%We can model this as binomial process, which returns 1 when the scalar is reconstructed correctly and returns 0 otherwise; the probability of returning 1 is $\epsilon$. Accordingly, the probability of correctly reconstructing $\z$ events out of $n$ is given by:
\begin{equation}
B(a|n, \epsilon) = {n \choose a} \epsilon^a(1-\epsilon)^{n-a}\label{Bdef}
\end{equation}
This distribution has mean $\epsilon n$ and standard deviation
\begin{equation}
 \sigma_n= \sqrt{n\epsilon (1 - \epsilon)} \label{varn}
\end{equation}

To describe the full process, where $n$ events are observed from an expected $N$ events and of that $n$, $a$ events are reconstructed correctly,
% by a jet algorithm of efficiency $\epsilon$, 
we combine the two probability distributions and sum over the intermediate variable $n$. For example, we can ask what is the probability of finding 5 events passing our cuts when we expect 100 to be produced? We have to sum over the probability of reconstructing 5 events from every possible value of the number of observed events, which can range from 5 to $\infty$. This can be expressed as:
\begin{equation}
P(a) = \sum_{n=a}^{\infty} \left[ P(n|N) \cdot B(a|n,\epsilon) \right]
\end{equation}
This distribution has mean $\epsilon N$, as expected, and variance $\sigma^2 =  N\epsilon$.
Thus the uncertainty in the number of background events measured is
\begin{equation}
 \delta B= \sqrt{N_B \epsilon_B} \label{deltaeN}
\end{equation}
The significance is then $S/\delta B = N_S \epsilon_S/\sqrt{N_B \epsilon_B}$.

%The expected number $\bar{a}$ of correctly reconstructed events is simply the expected number of events produced modified by the reconstruction efficiency to give $\epsilon N$. By taking the second moment of the above distribution we can show that the variance is also $\epsilon N$ which gives a characteristic uncertainty of $\sqrt{\epsilon N}$. 

In summary, the uncertainty associated with the number of events
 gets a contribution from the Poisson nature of the production process and another contribution from the uncertainty on whether an event will pass our cuts.
When both uncertainties are combined the mean and
 variance of the expected number are both $\epsilon N$.

%\subsection{Single event probabilities}
%\subsection{Statistics with Qanti-$k_T$ \label{sec:qstats}}
\subsection{Weighted cuts with Qjets \label{sec:qstats}}
A trivial observation which simplifies the uncertainty calculation for weighted events is that, since each event is independent, the probability that $a$ of $n$ events will pass a set of cuts is completely determined by the probability that one event will pass the cuts. This is true both for classical
algorithms which produces weights $\z = 0$ or $1$ and algorithms which combine multiple interpretations, like the pruned Qjets algorithm
used in~\cite{Ellis:2012sn} and Qanti-$k_T$ described here.
We start by rewriting the classical case calculation in terms of single event probabilities,
then discuss how the calculation is modified for weighted events. 
%To prepare for the uncertainty calculation using  Qjets, it is helpful to rewrite the observations about the classical case in suggestive notation. 

The cut-weight $\z$ denotes how signal-like a single event is: $\z=1$ is very much signal (by some measure) and $\z=0$ is very much background.
We can then define a  function $\f(\z)$ which gives the probability that an event passes the cuts. 
For the classical analysis, an event can only have $\z = 1$ (signal) or $\z = 0$ (background). 
Thus this probability function in the classical case is
\begin{equation}
\f_\text{class.} (\z) = (1- \epsilon) \delta(\z) + \epsilon \delta(\z-1) \label{fcdef}
\end{equation}
which  matches the classical anti-$k_T$ panel of Fig.~\ref{fig:f1s}.

What is the probability for a single event to pass a set of cuts?
We can compute this either using Eq.~\eqref{Bdef} with $a=0,1$ or with Eq.~\eqref{fcdef}
integrating over $\z$. The two methods agree:
\begin{equation}
\langle \z \rangle  =\int d\z \, \z\, \f_\text{class.}(\z)= \epsilon \label{mom1}
=
\sum_{a=0}^1 a B(a | n=1,\epsilon)
\end{equation}
Similarly, we find
\begin{equation}
\langle \z^2\rangle  =\int d\z \, \z^2\, \f_\text{class.}(\z)= \epsilon \label{mom2}
= \sum_{a=0}^1 a^2 B(a | n=1,\epsilon)
\end{equation}
Thus if we know that exactly one event is produced, we find
\begin{equation}
\sigma_{1,{\text{class}}}^2=\langle \z^2\rangle-\langle\z\rangle^2= \epsilon(1-\epsilon)
\end{equation}
as in Eq.\eqref{varn} with $n=1$.

To get the expected variance on the full distribution, we have to include the Poisson uncertainty
which depends on the mean $\langle \z \rangle = \epsilon$. By the central limit theorem,
since the events are uncorrelated,  the characteristic
uncertainty on the distribution where $N$ events are expected is
\begin{equation}
\delta_{\text{class}}=\sqrt{N\left(\sigma_{1,{\text{class}}}^2+\langle \z \rangle^2\right)}= \sqrt{N \epsilon}
\end{equation}
in agreement with Eq.~\eqref{deltaeN}. %We can check this formula in two limits. If 
%we know that exactly one event was produced, then $\delta_{\text{class}}=\sigma_{1,{\text{class}}}$ as
%expected. When there
%are no cuts applied, $\sigma_{1,{\text{class}}}=0$ and the uncertainty reduces to
%$\delta_{\text{class}}=\langle \z\rangle\sqrt{N} $, also as expected.

%So that the mean and the variance $\delta = \epsilon(1-\epsilon)$ 
%can be computed with $\f_\text{class.}$ instead of $B(a | n,\epsilon)$.

%When using Qjets, the Poisson process of producing the events remain unchanged. 
%
%Describing the reconstruction though is somewhat more complicated. Each event is interpreted multiple times and each interpretation is a slightly different reconstruction of the event. For a given event, a classical jet algorithm would return 0 or 1 depending on whether it was reconstructed correctly. If we were to histogram the outcomes over a large number of events we would then get a distribution peaked at 0 and at 1. In contrast for Qjets, we can ask what fraction of the interpretations of a given event reconstructed the event correctly. The result could then take any value between 0 and 1, and consequently the distribution would be continuous between 0 and 1.  See Fig.~\ref{fig:f1s} for an illustration of these distributions.

With cut-weights $0 \le \z \le 1$,
 would also like to know what the probability is that $a$ events pass our cuts if $n$ events were produced at the collider which was expected to produce $N$ events.
The new feature in the  Qanti-$k_T$ case is that $a$ and $\z$ are not necessarily integers.
With Qanti-$k_T$, each event is interpreted multiple times. For each event, a fraction $\z$ of the interpretations pass the cuts and $a$ is the sum of the $\z$ values over all the measured events.
In the 
Qanti-$k_T$ case the function $\f(\z)$ now has meaning for $0 \le \z \le 1$. 
Examples of $\f(\z)$ are shown for various $\alpha$ are shown in  Fig.~\ref{fig:f1s}. 

Although the different interpretations coming from Qjets for the same event are highly correlated,
each event is uncorrelated with any other. Thus, as with the classical case, the
probability of finding $a$ events satisfying the cuts when $n$ events are produced is
completely determined by the probability that one event will satisfy the cuts. 
That is, we do not need to know what the generalization of $B(a|n, \epsilon)$ is
in Eq.~\eqref{Bdef}, only that it is determined completely by $\f(\z)$.

We calculate the uncertainty with weighted events exactly as we did in the classical case with $\f(\z)$ replacing
$\f_\text{class.}(\z)$. That is, we calculate
%
%% The function  $\f(\z)$ is the generalization
%%%of Eq.~\eqref{fcdef} to the case where a single event can pass the cuts a fraction $\z$ of the time.
%%
%%Indeed,
%%these functions were shown in Fig. 
%%
%%
%%As with the classical case, t
%%The production of $n$ events is still described by a Poisson distribution with mean $N$. However, we must replace the binomial distribution describing a classical jet algorithm with an equivalent distribution describing Qjets. Rather than compute this function, we can use the central limit theorem, which
%%says that for a stochastic process
%%
%
%Classically, the efficiency of a jet algorithm is defined as the probability of correctly reconstructing a given event or the mean of $B(a|n,\epsilon)$. This can be computed from $\f_\text{class.}$ as in
%Eq.\eqref{mom1}. The analogous efficiency for Qjets for one event is:
\begin{equation}
\langle \z\rangle_\f = \int d\z \, \z\, \f (\z)
\end{equation}
and
\begin{equation}
\sigma_{1,\f}^2 \equiv  \left[ \int d\z \, \z^2\, \f (\z)\right]  - \langle \z\rangle_\f^2
\end{equation}
Then  if $N$  events should have been produced, the expected number to be observed is 
\begin{equation}
N_\text{expected}=  N \langle \z\rangle_\f 
\end{equation}
The uncertainty on this number is
\begin{equation}
\delta_{\text{Qjets}}=\sqrt{N\left(\sigma_{1,\f}^2+\langle \z \rangle_\f^2\right)} = \sqrt{N}\sqrt{ \langle \z^2 \rangle }
\end{equation}
which is just Poisson fluctuations multiplying the root-mean-square (RMS) of the distribution.

In summary, to use weighted events, instead of counting an event as either satisfying a set of cuts ($\z=1$) or not satisfying them ($\z = 0$), an
event can fractionally satisfy them, giving a weight $0 \le \z \le 1$. Then the number of observed events is the sum over these $\z$ values
over all events. For a signal process, this number is written as $S=N_S \langle \z \rangle_{\f_S}$ where $\f_S(z)$ is given by the cross
section for getting a $z$ value of a signal process, normalized to unit area and $N_S$ is the total number of signal events considered.
For background, $B=N_B \langle \z \rangle_{\f_B}$
 The characteristic size of fluctuations of $B$ is given, in the limit of large number of events where the
central-limit theorem can be applied, by $\delta B = \sqrt{ N _B\langle \z^2 \rangle_{\f_B} }$. So that
\begin{equation}
\text{significance} = \frac{N_S \langle \z \rangle_{\f_S}}{ \sqrt{ N_B \langle \z^2 \rangle_{\f_B} } } \label{sigofz}
\end{equation}
To see how much cut-weights can help, one can take the ratio of this value to the cut-based significance. The overall number of signal and background events considered, $N_S$ and $N_B$, conveniently drop out of such a ratio.

\subsection{Reweighting \label{sec:reweighting}}
The procedure we have described can be applied for any way of computing weights. Using multiple interpretations to generate the weight $z$ is natural and intuitive.
As a simple generalization, one can consider transforming the weight by any function $t(z)$ to see if significance can be improved. The optimal function will
be the one that produces an extremum of the functional
\begin{equation}
\text{significance}[t] \equiv \frac{ \langle t \rangle_{\f_S}}{ \sqrt{ \langle t^2 \rangle_{\f_B} } } 
= \frac{ \int_0^1 d z t(z) \rho_S(z) }{\sqrt{\int_0^1 d z [t(z)]^2 \rho_B(z)}}
\label{sigoft}
\end{equation}
The functional variation of the significance is
\begin{equation}
\frac{\delta \text{significance}[t]}{\delta t(z')} = \frac{\f_S(z')}{ \langle t^2 \rangle_{\f_B}^{1/2}} - \frac{t(z') \f_B(z') \langle t \rangle_{\f_S}}{ \langle t^2 \rangle_{\f_B}^{3/2} }
\end{equation}
This vanishes when
\begin{equation}
t(\z) = \frac{ \f_S(\z)}{\f_B(\z) }
\label{toff}
\end{equation}
up to an overall constant which has no effect on the significance enhancement.

%
%In the classical case, as  in Eq.~\eqref{mom2}, $\sigma_\text{class.} = \epsilon_\text{class.}$,
%but in the quantum case this is not guaranteed.
%
%% \begin{figure}
%% 	\centering
%% 	\includegraphics[width=0.48\textwidth]{f1q_4.pdf}
%% 	\includegraphics[width=0.48\textwidth]{f1q_6.pdf}
%% 	\includegraphics[width=0.48\textwidth]{f1c.pdf}
%% 	\includegraphics[width=0.48\textwidth]{f1q_7.pdf}
%% 	\caption{The distribution $f_1(a)$ for, from top left going clockwise, $\alpha=0.001$, $\alpha=0.1$, $\alpha=1$, and standard anti-$k_T$ for a sample of dijet background events (see Sec.~\ref{sec:demo}). \label{fig:f1s} }
%% \end{figure}
%
%Putting everything together, the probability of correctly reconstructing a events with Qjets is:
%
%\begin{equation}
%P(a) = \sum_{n=a}^{\infty} \left[ P(n|N) \cdot \f(n)\right]
%\end{equation}
%with mean $\epsilon_f N$. Again by taking the second moment we find that the variance is $N(\epsilon_f^2 + \sigma_1^2)$, where $\sigma_1^2$ is the variance of the $f_1(a)$ distribution. The uncertainty is then $\sqrt{N(\epsilon_f^2 + \sigma_1^2)}$.
%
%As before, the Poisson uncertainty associated with producing the events at a collider remains unchanged. However, the uncertainty stemming from Qjets now has the potential to be significantly smaller than the uncertainty from the classical algorithm. This is why the background uncertainty in Qjets can be smaller leading to an improvement in $S/\dB$. Of course, the signal uncertainty decreases as well and this has important consequences for exclusion where the signal uncertainty enters.
% 

To use these results in practice, suppose we are interested in how much luminosity it would take to see a certain signal over a certain background. We first compute the expected numbers
$N_S$  and $N_B$ of signal and background events produced at the collider for a given set of cuts. 
Given these cuts, we can calculate $\f_S(\z)$ and $\f_B(\z)$, as in Fig.~\ref{fig:f1s}. Thus
functions give us $\langle\z\rangle_{\f_S}$ and $ \langle\z^2\rangle_{\f_B}$ (as well as $\langle t \rangle_{\f_S}$ and $\langle t^2 \rangle_{\f_B}$ if we want to use reweighted events). 
% and $\delta_{1,\f_B}$.
We then calculate
$S=N_\text{expected}^\text{sig.}=N_S \langle\z\rangle_{\f_S}$
and $\dB=\sqrt{N_B \langle \z^2 \rangle_{\f_B}} $.
%and $\dB=\sqrt{N_B\left(\delta_{1,\f_B}^2+\langle \z \rangle_{\f_B}^2\right)}$.
The expected significance is given by $S/\dB$. 
 With data, one could just 
look for an excess over expected background. Then $S$ would be replaced by
$N_\text{observed} - N_\text{expected}^\text{bkg}$.
%
%Putting everything together we find that 
%\begin{equation}
%\label{eq:soverdelb}
%\frac{S}{\dB}\propto \frac{\epsilon_{fS}}{\sqrt{\epsilon_{fB}^2 + \sigma_{1B}^2}}
%\end{equation}
%where subscripts of $S$ and $B$ denote quantities measured on signal and background processes, respectively.  

 As a comparison of the cut-based, cut-weighted, and reweighed  approaches, we give the expected significance for each method in Table~\ref{tab:ZHcompare} for the $Z+H$ signal and $Z+b\bar{b}$ background samples.
Note that since  $S/\dB$ scales as $\sqrt{\cal L}$ (the square root of the luminosity), an improvement in $S/\dB$ of $28\%$ means that one can make measurements with a significance comparable to standard anti-$k_T$ using only $(\frac{1}{1.28})^2=61\%$ of the luminosity. On 
the other hand, since $S/\dB$ is proportional to
$N_S /\sqrt{N_B}$, for any $\f$ one can compare the significance
for different algorithms and cuts independent of the expected cross section and luminosity.

\begin{table}[t]
\centering
\begin{tabular}{|c|c|c|c|}
      \hline
	observables & cuts & cut-weighted & cut-reweighted\\
      \hline
	$m_{JJ}$ & 1.00 &  --  &  --  \\
\hline
	$\alpha=0$ & 1.00  & 0.79  &0.82\\
\hline
	$\alpha=0.1$ & 1.01 & 1.19  &  1.24 \\
\hline
	$\alpha=0.3$ & 1.00  & 1.22  & 1.28\\
\hline
	$\alpha=1.0$ & 1.02  & 1.18 & 1.24\\
%\hline
%	$\f(\z)_{\alpha=0.1}$ and  $m_{JJ}$ & 1.62 (-0.6\%) &  --  & 2.15 (+33\%)\\
\hline
\end{tabular}
    	\caption{Comparison of the significance using the cut-based, cut-weighted, and reweighted methods. 
The $m_{JJ}$ window used $110\gev < m_{JJ} < 140\gev$ taken from and the significance of this cut is normalized to 1.
The mass window has not been optimized (optimizing it on our samples leads to $104\gev < m_{JJ} < 136\gev$ and gives a significance of 1.03).
This same $110\gev < m_{JJ} < 140\gev$ window is used to compute the weight functions
$\rho(z)$ for signal and background. Cuts refers to the number $N_S/\sqrt{N_B}$  of events in a window, cutting on the $\f_S(\z)$ and $\f_B(\z)$ distributions
as well as $m_{JJ}$
in the Qanti-kT cases. ``Cut-weighted'' refs to using Eq.~\eqref{sigofz} and``cut-reweighted'' refers to using Eq. \eqref{sigoft} and~\eqref{toff}.
All numbers
are for the same $Z+H$ sample (signal) and $Z+b\bar{b}$ sample (background), as described in Sec.~\ref{sec:HZ}. 
%The numbers in parenthesis are the percent improvement over the significance of 1.56 resulting from a simple $m_{JJ}$ cut.
\label{tab:ZHcompare}}
\end{table}

\section{Example applications}
\label{sec:demo}

In this section we show how Qanti-$k_T$ can be useful for
a variety of searches. We will consider three signals, listed here in ascending order of complexity: (1) a  resonance decaying into dijets, (2) a resonance produced in association with a vector boson (including the $H+Z$ example), and (3)  pair production of two resonances. We will see that while Qanti-$k_T$ does little to improve ordinary dijet reconstruction, the significance  of more complex events can be improved by  50\% over a classical analysis.

For each event class, we first process the signal and background events with classical anti-$k_T$ at various different values of $R$. We then fix the value of $R$ which optimizes the $S/\dB$ ratio (which for classical anti-$k_T$ is simply $N_S \epsilon_S/\sqrt{N_B \epsilon_B}$).  We then use this value of $R$ in Qanti-$k_T$ and compute $S/\dB$ for different values of rigidity ($\alpha$). Qanti-$k_T$ is useful to the extent that $S/\dB$ is larger than $S/\dB$ for the classical analysis. How $S/\dB$ is computed in Qanti-$k_T$ was
discussed in the previous section.  Results are summarized in Table~\ref{tab:sdelb}.

\subsection {Simple resonance reconstruction}
We consider first a dijet resonance decay to gluons. The signal process is $pp\rightarrow \phi\rightarrow gg$ with $m_\phi=1~{\rm TeV}$. The background is dijet production in the standard model.
We consider the two hardest jets in each event, requiring both jets to satisfy $p_T(j)>425~{\rm GeV}$ and the diet mass to be  in the window $950~{\rm GeV}< m < 1050~{\rm GeV}$. 
For this process and these cuts, we find that $R = 1.1$ gives the best $S/\dB$ ratio using
classical anti-$k_T$

Running Qanti-$k_T$ on these samples, we find at most a 3\% improvement (see Table~\ref{tab:sdelb}). 
That we find only a small improvement  is perhaps not unexpected in this case.  With hard well-separated dijets, any algorithm and most jet sizes should be able to pick out the dijets and get
their invariant mass mostly correct. Since there is little ambiguity in the events' interpretation, there is little to gain from resampling with Qanti-$k_T$.

\begin{figure}
	\centering
	\includegraphics[width=0.45\textwidth]{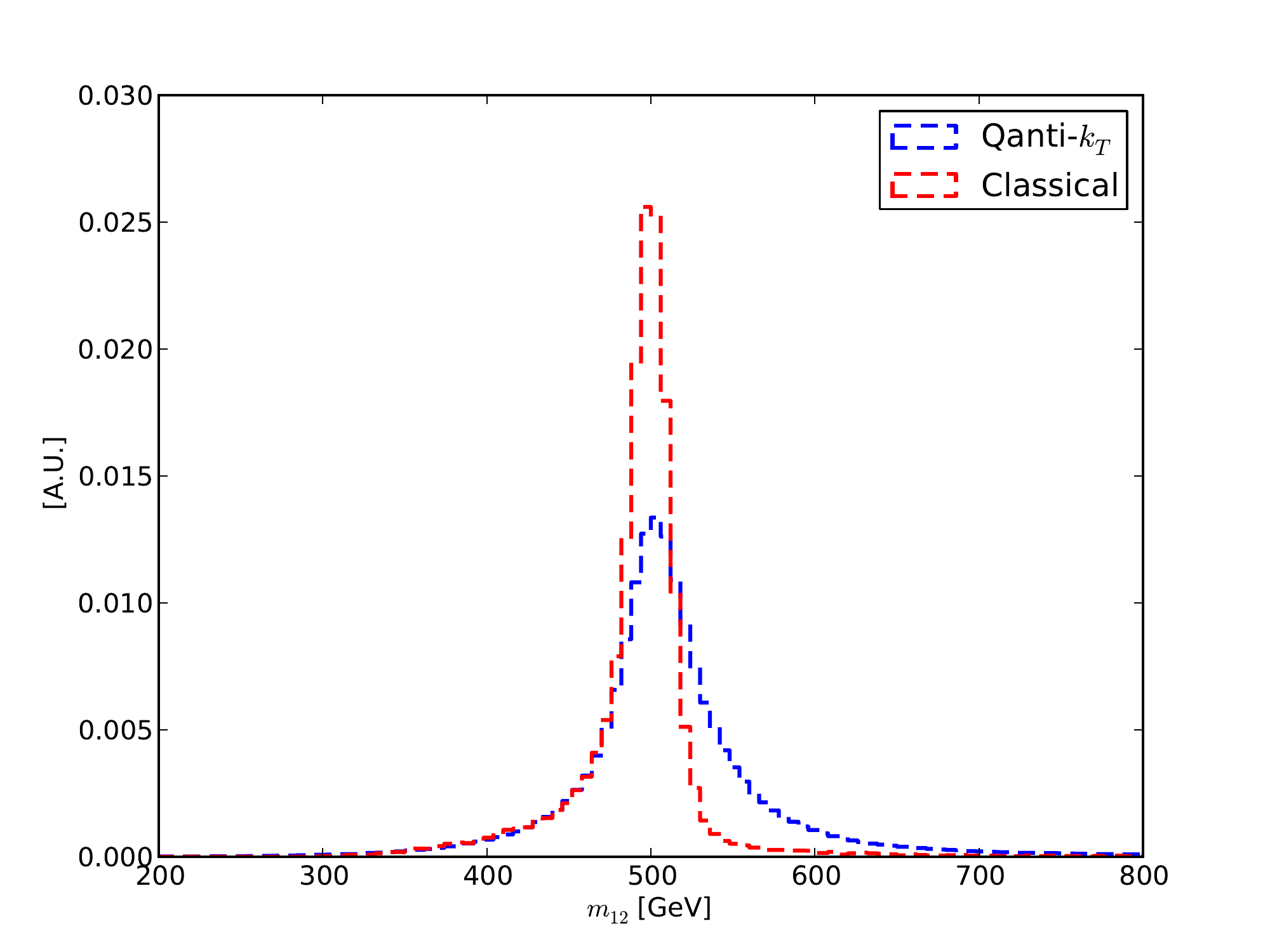}
	\includegraphics[width=0.45\textwidth]{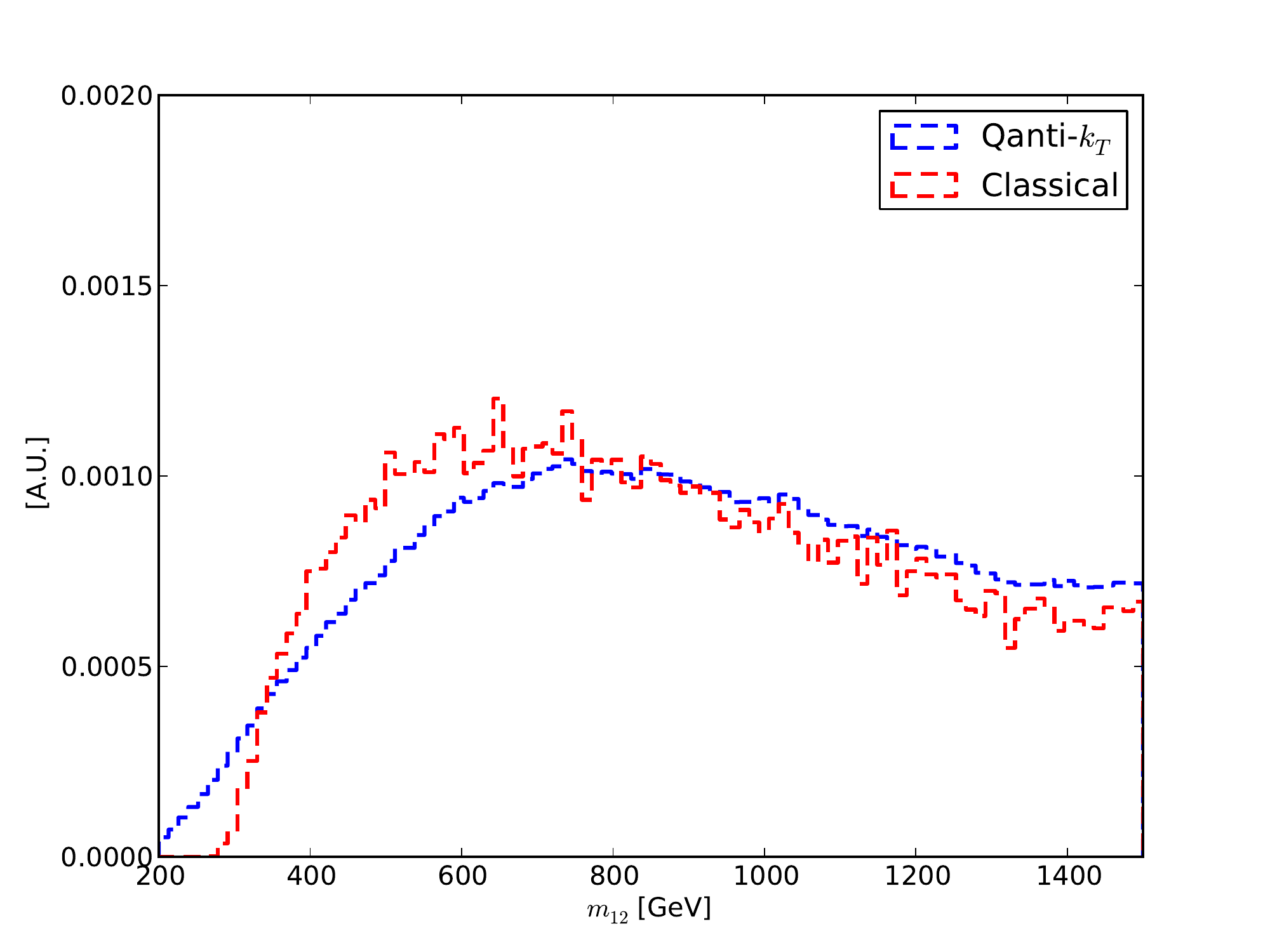}	
	\caption{A comparison of the signal (left) and background (right) dijet invariant mass distributions using standard anti-$k_T$ and Qanti-$k_T$ for optimized parameters. 
Signal is $Z\phi\to \nu \bar{\nu} gg$ with $m_\phi = 500\gev$ and background is $Zgg\to\nu \bar{\nu} gg$. All events have $\met>800 \gev$ and $p_T > 400 \gev$ for each jet. 
 \label{fig:dsigmadm} }
\end{figure}

\subsection{Boosted resonances in associated production}
Next we consider the case where a neutral scalar is produced in association with a $Z$ boson.  Unlike in the pure dijet case considered above, when the $Z$ boson and resonance have significant transverse momentum,
 the jets from the resonance decay will to be closer together and of unequal $p_T$. Thus, there will be more ambiguity about whether or not the jets pass the $p_T$ cut. For systems with a larger boost there will an additional ambiguity due to overlap between the jets.
 
First we consider the process $pp\rightarrow Z\phi\rightarrow gg\nu\bar\nu$ where $m_\phi=500~{\rm GeV}$.  The background is $pp \to Z+\text{dijets}$.
 We require  $\met>400~{\rm GeV}$, that the two hardest jets satisfy $p_T(j)> 200~{\rm GeV}$, and that the dijet invariant mass fall within the window $450~{\rm GeV}<m<550~{\rm GeV}$.
Here we find that the classical value of $R$ that optimizes $S/\dB$ is 0.95.  
Running Qanti-$k_T$ on this sample, we find a 9\% improvement in $S/\dB$ at $\alpha = 1.0$.

 That the improvement is larger in this case than without the boost is consistent with the intuition that 
Qanti-$k_T$ helps more when the interpretation of an event is more ambiguous.
For boosted resonances, the jet boundaries are close together. A classical algorithm, which only takes one interpretation of the event could easily assign radiation to the wrong jet. 
With 100 different interpretations of an  event,  some fraction of those interpretations will more correctly reconstruct the two jets than the classical algorithm
%\subsubsection{Increasing the boost}

Considering the same $500\gev$ scalar but going to higher boost, Qanti-$k_T$ helps even more. 
We next require $\met>800~{\rm GeV}$ and $p_T > 400$ GeV for each of the jets. This selects the events where the jets are even closer together. Here the  optimal $R$ value is found to be 0.65. 
Using this value of $R$, we find that with $\alpha = 0.1$, Qanti-$k_T$ produces a $S/\dB$ 
 19\% larger than in the classical case.  
 
We show the distribution of $m_{JJ}$ for signal and background for the classical and Qanti-$k_T$
samples in Fig.~\ref{fig:dsigmadm}.
In the classical case, each event contributes a single value of $m_{JJ}$. For Qanti-$k_T$, 
each event contributes many (100 in our samples) values of $m_{JJ}$.
 Although the Qanti-$k_T$ mass peak is broader for signal 
(so that $S$ goes down)  the improvement in the background stability (so that $\dB$ goes down)
provides sufficient compensation so that $S/\dB$ goes up overall.

\subsection{Higgs +Z \label{sec:HZ}} 
The boosted resonance analysis can be applied to Higgs boson production. 
Although boosted Higgs production can be considered with  jet substructure methods
\cite{Butterworth:2008iy}, these methods require the boost to be so large that the Higgs
decay products
merge into a single fat jet. For Qanti-$k_T$, the boost does not have to be so extreme. 
In fact, unlike substructure techniques, Qanti-$k_T$ will never degrade significance  (although it sometimes will not help much) as long as $\alpha$ is optimized, since $\alpha=\infty$ reduces
to the classical case.

We consider $H+Z$ production where the $Z$ decays to neutrinos and the Higgs decays to a $b$-quark pair  $(ZH \rightarrow \nu \bar{\nu} b \bar{b})$. As background, we take $Z+b\bar{b}$ production
without a Higgs.
We require that events yield at least two jets with $p_T > 25$ GeV, $\met>120~{\rm GeV}$, and that the invariant mass of the hardest two jets fall within the window $110\gev<m<140\gev$.  
The optimal $R$ value for the classical analysis in this case is 0.7.
Taking $R=0.7$ we find that with $\alpha$ is optimized with $\alpha = 0.3$ the $S/\dB$ improves by 22\% using the weighted-cuts approach and
by 28\% if we reweight by $\rho_S/\rho_B$ as discussed in Sec.~\ref{sec:reweighting} (see also Table~\ref{tab:ZHcompare}).

28\% is a substantial improvement in significance for an $H\to b\bar{b}$ channel. Indeed, a classical multivariate approach involving a sinkful of kinematic and substructure variables~\cite{Gallicchio:2010dq} was only able to achieve improvements of significance of order 20\%. 
Moreover, the $p_T$ cut of $120\gev$ (which can easily be lowered) is not
as extreme as the $200\gev$ cut proposed in~\cite{Butterworth:2008iy}, thus more signal events
can enter the Qanti-$k_T$ analysis than the boosted one. This at least suggests that the multiple-interpretations approach warrants more detailed study for Higgs searches.

\subsection{Resonance pair reconstruction \label{sec:fourjets}}

\begin{figure}
	\centering
	\includegraphics[width=0.48\textwidth]{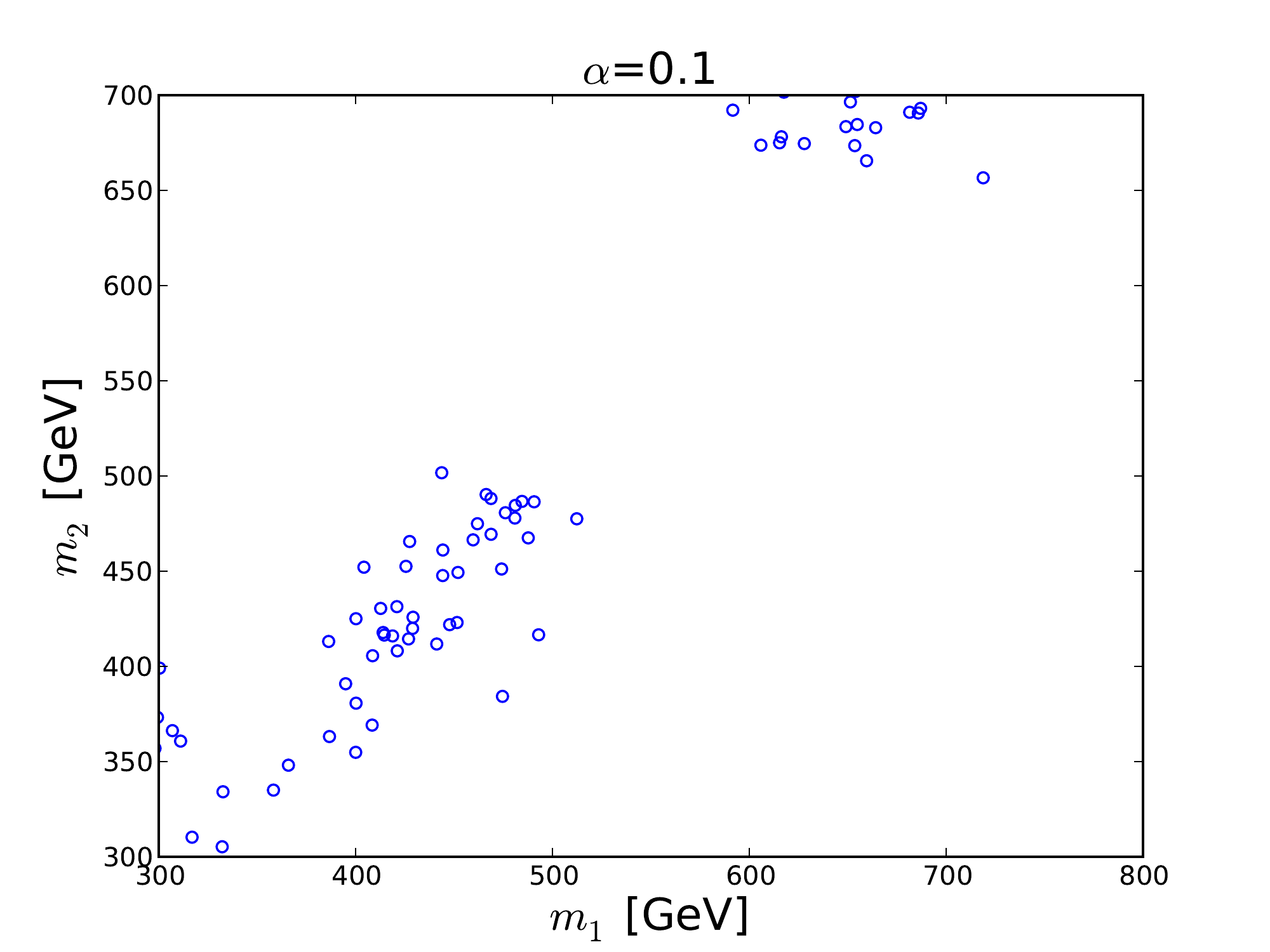}
	\includegraphics[width=0.48\textwidth]{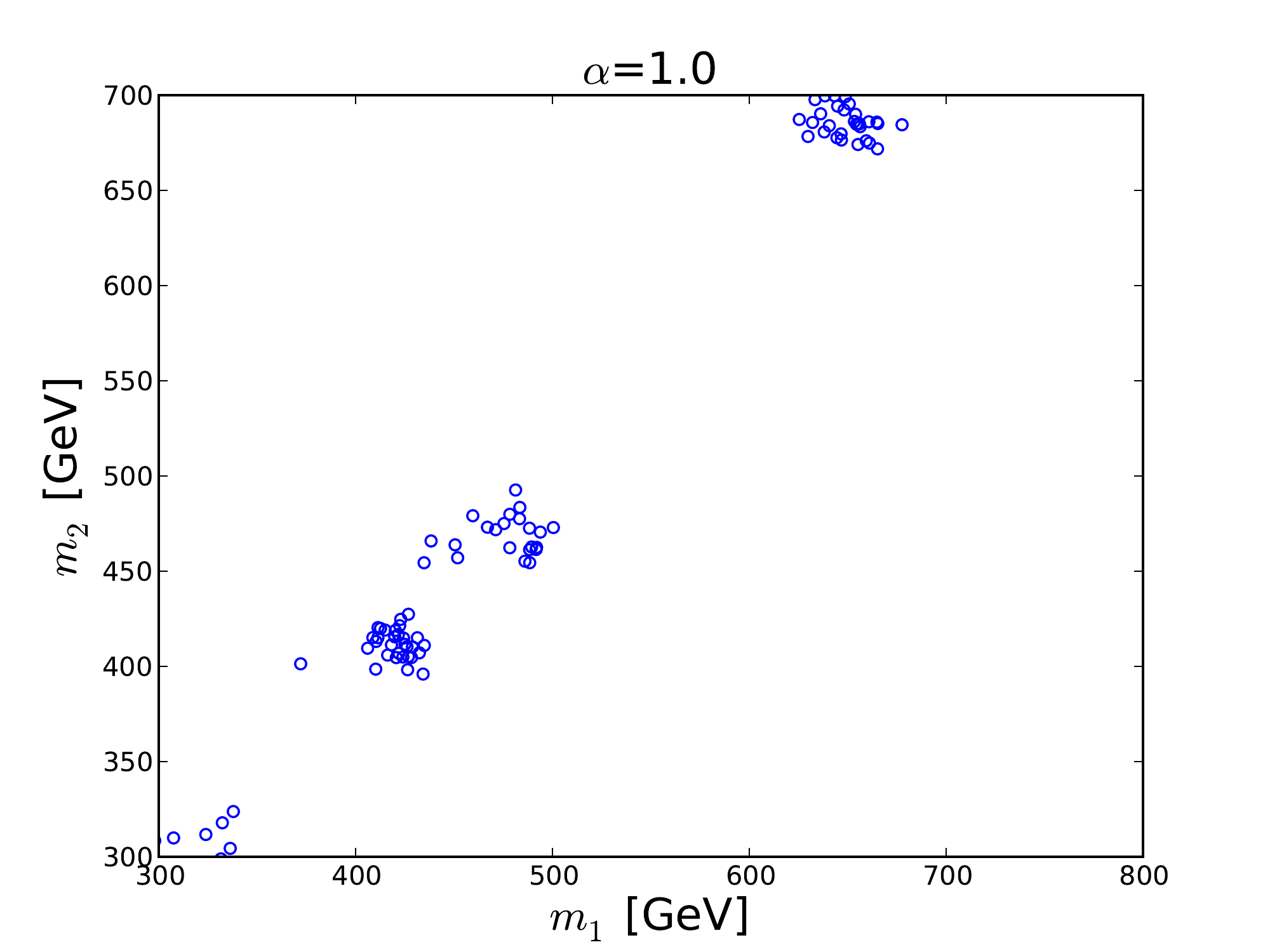}
	\includegraphics[width=0.48\textwidth]{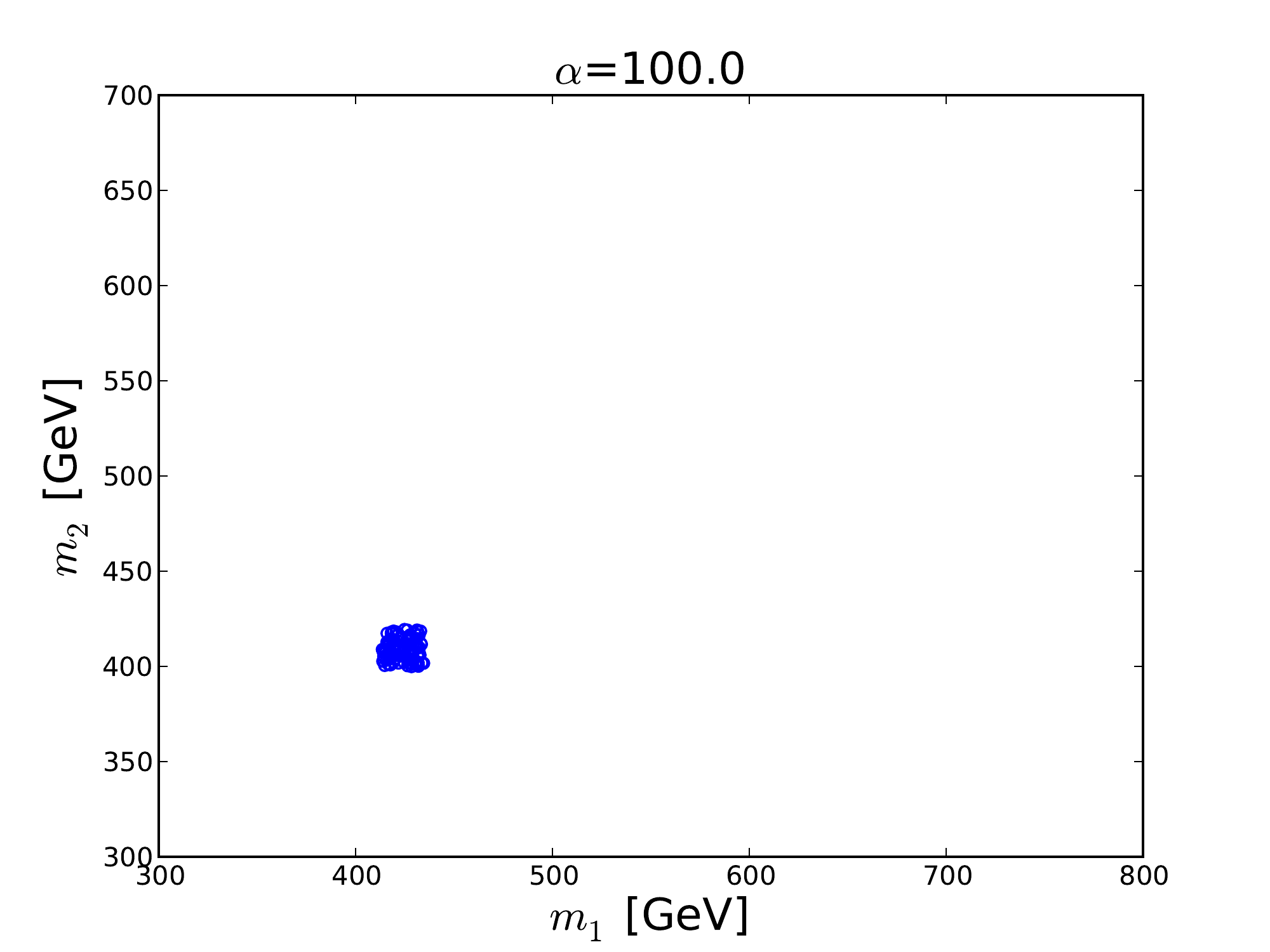}
	\includegraphics[width=0.48\textwidth]{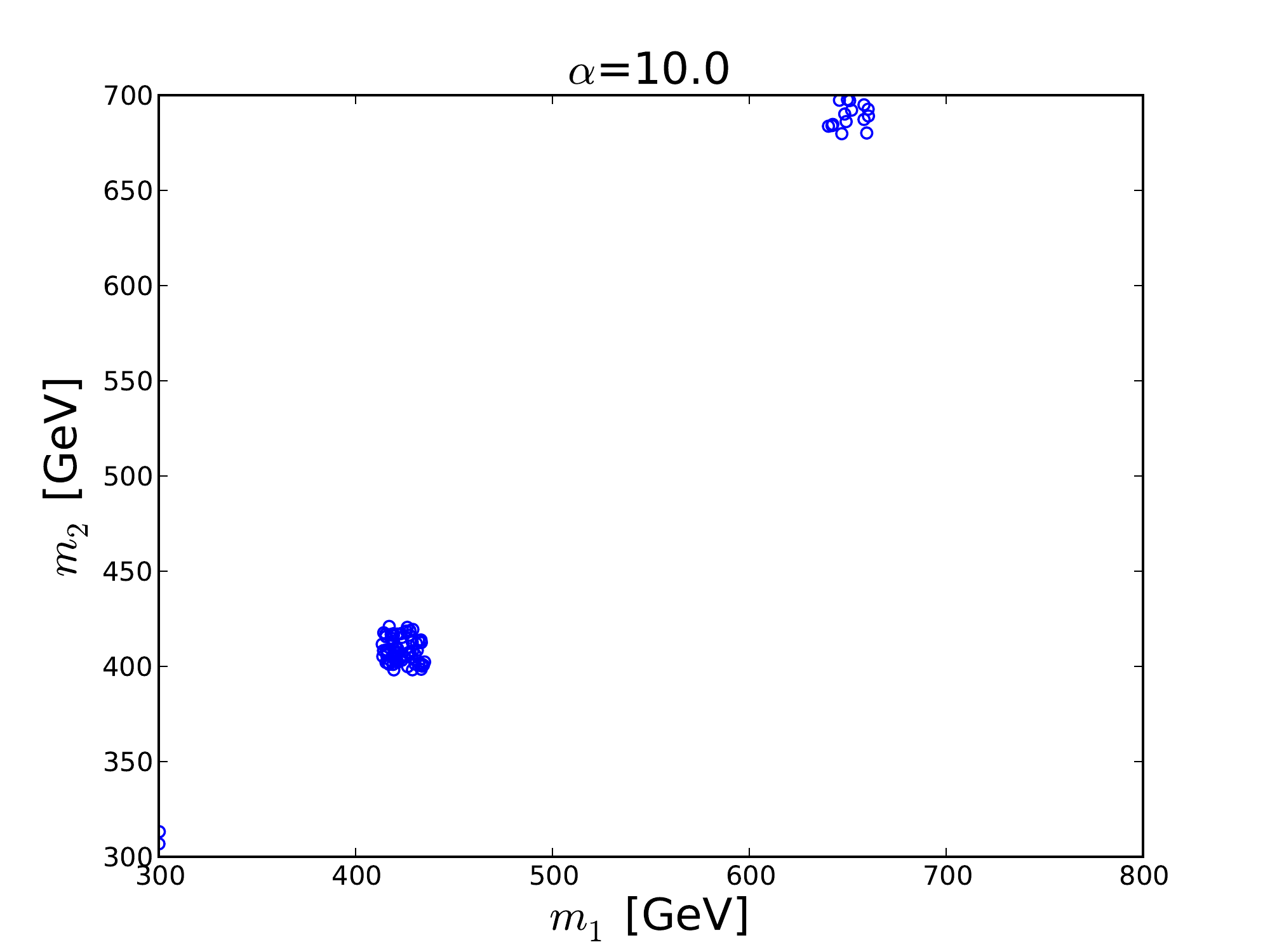}

	\caption{The two resonance masses in the $pp\rightarrow \phi \phi$ process found for 100 interpretations of a single signal event using Qjets. From top left going clockwise, $\alpha=0.1$, $\alpha=1$, $\alpha=10$, and $\alpha=100$.  We see that while the $\alpha\rightarrow\infty$ interpretation of the event does not fall within the mass window, such an interpretation arises when $\alpha$ is relaxed to $\sim1$ and below. \label{fig:respair}}
\end{figure}

Next, we consider four-jet events to test how well Qanti-$k_T$ works in a more complex jet environment. We consider the process $pp\rightarrow \phi \phi$, $\phi \rightarrow gg$, where $\phi$ is again a neutral scalar with  $m_\phi=500~{\rm GeV}$ (see Ref.~\cite{ATLAS:2012ds,Chatrchyan:2013izb} for similar analyses at ATLAS and CMS). The background in this case is four-jet production in QCD. 
%In this case, we  expect even more ambiguity than in the two-jet cases we have considered. 

In analyzing the four-jet events, while the core Qanti-$k_T$ algorithm remains unaltered we add a preselection step to speed up the analysis (cf.  Sec.~\ref{sec:speed}). In the preselection step we run both signal and background events through anti-$k_T$ using Fastjet with $R = 0.5$. We then check to see if each of the four hardest jets in each event have $p_T > 120$ GeV. Only events passing this cut are passed through to our non-deterministic anti-$k_T$ algorithm.

Each interpretation of each event using Qanti-$k_T$ (or the single classical interpretation) gives a set of jets.
Our goal is to select from these jets the four that yield two pairs which are close to each other in mass. In order to do this, we select the five hardest jets from the final set of jets, form all possible pairs, and calculate the invariant mass for each pair. For each two pairs $a$ and $b$ (representing the reconstructed scalars),  we calculate the quantity $|m_{a} - m_{b}|/(m_{a} + m_{b})$ to evaluate how close in mass the reconstructed scalars are. We choose the pairing that minimizes the mass difference between the two reconstructed scalars. Once the pairing is chosen, we further require that:
\begin{itemize}
\item The mass difference between the two reconstructed scalars is less than 20\%:  $|m_{a} - m_{b}|/(m_{a} + m_{b}) < 0.2$
\item The  average mass $(m_{a} + m_{b})/2$ of the two reconstructed scalars fall within the window $450 - 550 \gev$.
\item Each jet used to reconstruct the scalars must have $p_T > 120$ GeV
\end{itemize}

An example distribution of $m_{a}$ vs $m_{b}$ for a single event is shown in Fig.~\ref{fig:respair}. We see that the classical analysis ($\alpha \sim 100$) does not find $m_a=m_b=500 \gev$ which would correspond to perfect reconstruction. The distribution of $m_a$ and $m_b$ for finite $\alpha$ shows
that many masses can be sampled. More importantly, we see that some samplings come very close to the perfect reconstruction. This shows why Qanti-$k_T$ will be helpful for this multijet sample.

This procedure is applied first to the classical analysis. We find that
$R = 0.75$ maximizes $S/\dB$ in the classical case.
Using this value of $R$, the $S/\dB$ improvements using Qanti-$k_T$ on the same signal and background events at different values of the rigidity parameter $\alpha$ are shown in Table~\ref{tab:sdelb}. We see that at $\alpha = 0.1$ there is a 49\% improvement in $S/\dB$ over the classical results. 
As with the previous cases, when $\alpha$ approaches higher values such of 10 and 100, the improvement declines as the algorithm begins to behave more like the classical algorithm. At very low values of $\alpha$ the performance of Qanti-$k_T$ is poor. Again this is expected due to the highly random nature of the mergings at low $\alpha$ with little physical motivation

The large improvement (49\%) in significance achievable with Qanti-$k_T$ over the classical analysis is consistent with our expectation that Qanti-$k_T$ helps more in more complicated event topologies. In this case, having four jets rather than two makes the jets more likely to overlap and Qanti-$k_T$ is more likely to be helpful.

\begin{table}
\centering
  \begin{tabular}{ |c | c | c | c | c | c | c | c | c | c | c|}
    \hline
	\multirow{2}{*}{Sample}& \multirow{2}{*}{R} &\multicolumn{6}{c|}{Improvement in $S/\dB$ (\%)} \\
%\multicolumn{6}{c|}{\small $\epsilon_{fS}$/ $\epsilon_{fB}$/ $\sqrt{\sigma^2_{1B}}$} 
	& & anti-$k_T$  &$\alpha=0$ &$\alpha=0.01$&  $\alpha=0.1$& $\alpha=1$& $\alpha=100$\\
	\hline
	\hline	\multirow{1}{*}{$pp\rightarrow \phi$}&\multirow{1}{*}{1.10}& 1.0&0.14&0.77&0.89&{\bf 1.03}&1.01\\
%	& & {\tiny 0.26/0.06/0.24}&{\tiny 0.01/0.01/0.05}&{\tiny 0.08/0.03/0.10}&{\tiny 0.16/0.05/0.16}&{\tiny 0.22/0.06/0.19}&{\tiny 0.25/0.06/0.23}\\
	\hline
	\multirow{1}{*}{$pp\rightarrow \phi+Z$(A)}&\multirow{1}{*}{0.95} &1.0&0.64&0.99&1.07&{\bf 1.09}&1.01\\
%	& & {\tiny 0.50/0.06/0.23}&{\tiny 0.26/0.06/0.24}&{\tiny 0.26/0.06/0.24}&{\tiny 0.26/0.06/0.24}&{\tiny 0.26/0.06/0.24}&{\tiny 0.26/0.06/0.24}\\
	\hline	\multirow{1}{*}{$pp\rightarrow \phi +Z$(B)}&\multirow{1}{*}{0.65}&1.0&0.58&0.98&{\bf 1.19}&1.10&1.01\\
%	& & {\tiny 0.26/0.06/0.24}&{\tiny 0.26/0.06/0.24}&{\tiny 0.26/0.06/0.24}&{\tiny 0.26/0.06/0.24}&{\tiny 0.26/0.06/0.24}&{\tiny 0.26/0.06/0.24}\\
	%& & {\tiny 0.26/0.06/0.24}&{\tiny 0.26/0.06/0.24}&{\tiny 0.26/0.06/0.24}&{\tiny 0.26/0.06/0.24}&{\tiny 0.26/0.06/0.24}&{\tiny 0.26/0.06/0.24}\\
	\hline	\multirow{1}{*}{$pp\rightarrow h +Z$}&\multirow{1}{*}{0.7}&1.0&0.79&0.99&{\bf 1.19}& 1.18&1.01\\
	%& & {\tiny 0.26/0.06/0.24}&{\tiny 0.26/0.06/0.24}&{\tiny 0.26/0.06/0.24}&{\tiny 0.26/0.06/0.24}&{\tiny 0.26/0.06/0.24}&{\tiny 0.26/0.06/0.24}\\
	\hline	\multirow{1}{*}{$pp\rightarrow \phi +\phi$}&\multirow{1}{*}{0.75}&1.0&0.75&1.43&{\bf 1.49}&1.40&1.01\\
    \hline
  \end{tabular}
    	\caption{The improvement in $S/\dB$ compared to standard anti-$k_T$ for various processes using different values of $\alpha$, the rigidity parameter.  $pp\rightarrow \phi +Z$(A/B) denote the $\phi+Z$ processes with a missing energy cuts of $400$ and $800~{\rm GeV}$, respectively. The value of $R$ used in both standard anti-$k_T$ and Qanti-$k_T$ is the one which optimizes the standard anti-$k_T$ results.  The largest improvements are shown in {\bf bold}. \label{tab:sdelb}}
\end{table}

\section{Speed}
\label{sec:speed}
Unfortunately, adding non-determinism to a jet algorithm and running it 100 times 
can slow down an analysis significantly. You might expect that running something 100 times (with no optimization)
 should take at worst 100 times the amount of time it takes to run it once. But actually,
 our algorithm can be even slower. The reason  is that one must recompute $\omega_{ij}^{(\alpha)}$ at each stage in the clustering using a new $d^{\rm min}$ (see Eq.~\ref{eq:omega}), whereas ordinary anti-$k_T$ need only compute the smallest distance at each step. 
Because of this extra information required, we cannot exploit without modification  the computational geometry techniques~\cite{Cacciari:2005hq} which makes fastjet fast.
The result is that it can take  tens of seconds per CPU to run 100 iterations on a event
with   several hundred particles. This is more
of an inconvenience than a problem at the current time. Nevertheless, it would be nice
to speed Qanti-$k_T$ up.

While our unoptimized  implementation (available at  \url{http://jets.physics.harvard.edu/Qantikt})
 is fast enough for practical use there are a few methods one can employ to speed it up.  These include:
\begin{itemize}
\item {\bf Preselection}:  To avoid unnecessary computation it can be helpful to first require all events pass a loose set of cuts using classical anti-$k_T$ jet before running anti-$k_T$ non-deterministically.
  This can significantly reduce the number of events processed.  For instance, if one is interested in a computation of dijet invariant mass for all events with satisfying $p_T\geq 100~{\rm GeV}$ one might first apply a preselection cut requiring all events have classical 
  anti-$k_T$ jets which satisfy $p_T\geq 75~{\rm GeV}$.  
\item {\bf Limited mergings}:  Rather than computing the distance between each pair of four-momenta one can make the physically motivated assumption that a pair of particles further apart than some large distance (say, $\Delta R \geq 2.0$) are unlikely to be part of the same jet.  Such pairings can be excluded from the analysis to improve the algorithm execution time.
\item {\bf Preclustering}: The runtime scales as $n^2\ln n$ for $n$ the number of particles to cluster, so its performance is quite sensitive to the number of initial particles.  An easy way to reduce the number of particles used as input to the algorithm is to first cluster them into larger micro jets or into a coarser grid.  For instance, if one finds that jets of $\delta \phi\times\delta\eta=0.1\times 0.1$ yield an algorithm which is too slow, one can merge these into $\delta \phi\times\delta\eta=0.2\times 0.2$ cells to realize a ${\cal O}(16\times)$ speedup.  
\item {\bf Optimization}: Since much of the distance information is reused from iteration to iteration,
there is plenty of potential to speed up the analysis by not recomputing these distances each iteration. 
More generally, smarter programming should speed up the algorithm significantly, as in fastjet~\cite{Cacciari:2005hq}. 
\item {\bf Modification}: Our non-deterministic anti-$k_T$ algorithm is in a sense the simplest
way to apply Qjets at the event level. One can easily conceive of other methods
which might be better suited to speed-up. A promising approach which just clusters once and then varies the jet size $R$
is discussed in~\cite{YT}.
\end{itemize}

\section {Conclusion}
\label{sec:conc}
In this paper, we have presented a fundamentally new way to think about events with jets.
Traditional algorithms, such as anti-$k_T$, give a single interpretation of an event. This
interpretation can be thought of as a best guess at the assignment of particles into jets.
These jets are meant to represent which particles came from the showering and fragmentation of which hard particle.
 In many events, however, there can be significant ambiguities
in which particle belongs to which jet. These ambiguities show up, for example,
in how different jet algorithms or jet sizes can give vastly different results for infrared
safe observables. The problem is that each algorithm gives a single best guess no matter
what -- ambiguous events and unambiguous events are treated the same way,
and  all information about the ambiguity is lost. 
In other words, an event which is clearly signal-like by some measure is given the same influence over the results as an event which is marginally signal-like (in the sense that it
would no longer be signal-like under a small change of parameters).  

The idea behind Qjets, which we have used here on the the event level, is that the ambiguity 
provides useful information about an event. By making a jet algorithm non-deterministic, we can compute the distribution of interpretations around the classical interpretation via Monte-Carlo sampling. When a non-deterministic jet algorithm (for example the Qanti-$k_T$ algorithm we
present here) is run 100 times on an event the event 100 different interpretations result.
The larger the variation in these interpretations, the more ambiguous an event is.
We introduce a parameter  $\alpha$, called rigidity after a similar parameter in~\cite{Ellis:2012sn},
which interpolates between classical anti-$k_T$ ($\alpha=\infty$) and purely random clustering ($\alpha=0$).

There are many ways that an ensemble of interpretations  can be used. The simplest way is to construct a $Q$-observable such as the variance of some classical observable (like the jet mass) over the interpretations. An example of this approach is the volatility variable introduced in~\cite{Ellis:2012sn}. One can then cut on this variance to improve significance. However, since almost all events are signal-like to some extent, it makes more sense
to include all the events in the analysis, with a weight based on the fraction interpretations passing a set of cuts.
 We derive a formula for the significance using weighted events which can be used to incorporate information from all
 the interpretations of all the events, rather than cutting some events out all together.

We applied Qanti-$k_T$ to a number of types of events.
We find that unambiguous processes, like those which produce hard and well-separated jets, do not benefit much from this procedure. However, for  more complicated processes, such as those with softer or overlapping jets, the significance can be improved significantly.
In a toy example, we showed that pair production of dijet resonances one can realize a $49\%$ improvement in $S/\dB$. 

Using weighted events from multiple interpretations has the potential to  improve substantially searches for the Higgs boson and measurements of its properties. We found that  for $pp \to ZH \to \nu \bar{\nu}b\bar{b} $ events 
at the 8 TeV LHC with $p_T^Z > 120\gev$, one can realize an 28\% improvement in significance over an equivalent classical analysis. We chose this $p_T$ fairly arbitrarily. With a $p_T$ cut less than
$120~\gev$ and we still expect Qanti-$k_T$ to improve significance, although perhaps not by as much.
That is, the methodology of using multiple interpretations not restricted to the highly boosted regime, as are other approaches
to finding the Higgs in this channel~\cite{Butterworth:2008iy}.
For other Higgs associated production channels (such as $pp \to WH$ and $pp \to t\bar{t} H$)
with $H\to b\bar{b}$, we expect the Qjets framework to be similarly helpful.

The Qanti-$k_T$ algorithm introduced in this paper be used whenever ordinary anti-$k_T$ is employed. 
While more complex event topologies tend to benefit more, Qanti-$k_T$ will at least never make an analysis worse. Indeed, since  for $\alpha=\infty$, Qanti-$k_T$ reduces to ordinary anti-$k_T$, as long
as one scans over $\alpha$, no harm can come (other than wasting time). It is natural to consider
applying Qanti-$k_T$ or some variation within the Qjets framework to challenging processes, such as top-tagging. When tops are very boosted,
it is likely that substructure methods will work better~\cite{Kaplan:2008ie} (although merging
Qjets with substructure is also promising), however, in the intermediate regime~\cite{Plehn:2009rk}
with moderate boost Qjets could help a lot.
It would also be interested to see if Qjets can help with color flow~\cite{Gallicchio:2010sw, Hook:2011cq},  quark and gluon discrimination~\cite{Gallicchio:2011xq,Gallicchio:2012ez}, ISR tagging~\cite{Krohn:2011zp} or in any situation where
ambiguities in reconstruction are problematic.

\acknowledgments

The authors would like to thank Yang-Ting Chien, Steve Ellis, David Farhi, Andrew Hornig, Cristina Popa, Lisa Randall, Keith Rehermann, and Tuhin Roy for useful discussions.  D. Kahawala is supported by the General Sir John Monash Award, D. Krohn is supported by a Simons postdoctoral fellowship and by an LHC-TI travel grant, and   MDS is supported by DOE grant DE-SC003916. Our computations were performed on the Odyssey cluster at Harvard University.

\bibliography{bio}
\bibliographystyle{jhep}
\end{document}